\documentclass[%
 aip,
 amsmath,amssymb,
preprint,%
]{revtex4-1}
\usepackage{subcaption}
\usepackage{graphicx}
\usepackage{cleveref}
\usepackage{epstopdf}
\Crefname{figure}{Figure}{Figures}
\crefname{figure}{figure}{figures}
\Crefname{table}{Table}{Tables}
\crefname{table}{table}{tables}
\usepackage{soul, color, xcolor}
\usepackage{dcolumn}
\usepackage{bm}
\usepackage{makecell}
\usepackage{multirow}
\usepackage[utf8]{inputenc}
\usepackage[T1]{fontenc}
\usepackage{mathptmx}
\usepackage{booktabs}
\usepackage{etoolbox}
\usepackage{array}
\makeatletter
\def\@email#1#2{%
 \endgroup
 \patchcmd{\titleblock@produce}
  {\frontmatter@RRAPformat}
  {\frontmatter@RRAPformat{\produce@RRAP{*#1\href{mailto:#2}{#2}}}\frontmatter@RRAPformat}
  {}{}
}%
\makeatother
\begin{document}

\preprint{AIP/123-QED}

\title[Unsteady flow predictions around an obstacle
using GP-DE-PINN]{Unsteady flow predictions around an obstacle
using Geometry-Parameterized Dual-Encoder Physics-Informed Neural Network}
\author{Zekun Wang}

\author{Yu Yang}%
\altaffiliation{Corresponding author, Email: yuyangyy@sjtu.edu.cn}

\affiliation{ 
Marine Numerical Experimental Center, State Key Laboratory of Ocean Engineering, School of Ocean and Civil Engineering, Shanghai Jiao Tong University, Shanghai 200240, PR China}

\author{Linyuan Che}
\affiliation{State Key Laboratory of Maritime Technology and Safety, 
Shanghai Ship and Shipping Research Institute Co., Ltd., 
Shanghai 200131, PR China}

 \author{Jing Li}%
\affiliation{ 
Marine Numerical Experimental Center, State Key Laboratory of Ocean Engineering, School of Ocean and Civil Engineering, Shanghai Jiao Tong University, Shanghai 200240, PR China}

\date{\today}

\begin{abstract}

Machine learning-based flow field prediction is emerging as a promising alternative to traditional Computational Fluid Dynamics, offering significant computational efficiency advantage. In this work, we propose the Geometry-Parameterized Dual-Encoder Physics-Informed Neural Network (GP-DE-PINN) with a dual-encoder architecture for effective prediction of unsteady flow fields around parameterized geometries. This framework integrates a geometric parameter encoder to map low-dimensional shape parameters to high-dimensional latent features, coupled with a spatiotemporal coordinate encoder, and is trained under the Navier-Stokes equation constraints. Using 2D unsteady flow past petal-shaped cylinders as an example, we evaluate the model's reconstruction performance, generalization capability, and hyperparameter sensitivity. Results demonstrate that the GP-DE-PINN significantly outperforms the PINN with direct geometric input in flow field reconstruction, accurately capturing vortex shedding structures and pressure evolution, while exhibiting superior generalization accuracy on unseen geometric configurations. Furthermore, sensitivity analyses regarding geometric sampling and network width reveal the model's robustness to these hyperparameter variations. These findings illustrate that the proposed framework can serve as a robust and promising framework for predicting unsteady flows around complex geometric obstacles.

\end{abstract}

\maketitle

\section{\label{sec:level1}Introduction}

Understanding the interactions between fluid flows and solid structures is fundamental for optimizing engineering systems, such as offshore platforms and aerospace vehicles \cite{afridi2024fluid, hou2012numerical}. In these scenarios, predicting the flow dynamics around obstacles with varying geometries is a common task, where geometric data determine the flow state. However, the application of high-fidelity Computational Fluid Dynamics (CFD) simulations in multi-query optimization tasks faces bottlenecks: generating high-quality body-fitted meshes for complex shapes requires specialized expertise \cite{karniadakis2021physics, wang2013high}, and resolving multi-scale spatiotemporal features consumes substantial computational resources \cite{kochkov2021machine, vinuesa2022enhancing}. These constraints have motivated research into accelerated computational strategies aimed at bypassing mesh generation while maintaining predictive accuracy \cite{brunton2020machine, tao2024multi}.

The integration of deep learning and computational mechanics offers a new approach to addressing these efficiency barriers \cite{YANG2023113470,YANG2025379,ZHAO2025122061,10.1063/5.0301145}. Early surrogate models, such as Convolutional Neural Networks (CNNs), demonstrated the capability to approximate flow fields faster than numerical solvers by learning mappings from geometry to flow variables \cite{guo2016convolutional, bhatnagar2019}. However, these data-driven models are based on large-scale high-fidelity simulation datasets and often do not satisfy conservation laws when extrapolating to unseen regimes \cite{willard2022integrating}. Physics-Informed Neural Networks (PINNs) have been proposed as an alternative framework \cite{raissi2019physics}. By embedding the Navier-Stokes equations into the loss function, PINNs can operate as mesh-free solvers \cite{lu2021deepxde, cuomo2022scientific, shah2024physics}. This method based on physics laws applies to the solution of forward and inverse problems ranging from laminar flows to turbulent systems \cite{cai2021physics}.

There have been numerous studies utilizing PINNs for flow field reconstruction. Pioneering work in this domain was established by Raissi et al., who introduced the concept of "hidden fluid mechanics". They demonstrated the ability to infer full-field velocity and pressure distributions solely from observed scalar dye concentrations, thereby bridging the gap between qualitative visualization and quantitative measurement \cite{raissi2020hidden}. Building on this foundation, subsequent research has explored the potential of recovering flow fields from sparse sensor data. For instance, Jin et al. systematically validated the reconstruction capabilities of PINNs in both incompressible laminar and turbulent flows, demonstrating their robustness in solving inverse problems \cite{jin2021nsfnets}. Specifically addressing the classic problem of flow past a cylinder, Xu et al. \cite{xu2021explore} proposed a physics-informed deep learning framework that successfully reconstructed the wake flow field from sparse velocity observations. Their work accurately identified missing parameterized dynamics within the governing equations, validating the method's high precision in capturing vortex shedding structures \cite{ren2024physics, chaurasia2024reconstruction}. Collectively, these studies indicate that PINNs have emerged as a powerful tool for bridging sparse observational data with the fundamental laws of fluid physics.

Currently, some researchers are no longer limited to using PINNs to reconstruct a single flow field. Integrating geometric information with flow prediction networks is a current focus in scientific machine learning. Sun et al. \cite{sun2023physics} utilized parameterized shape coefficients as inputs for a PINN to perform simultaneous surrogate modeling and optimization of airfoils, enabling design iterations without adjoint codes. Addressing irregular or discrete geometric representations, Kashefi and Mukerji \cite{kashefi2022physics} proposed a Physics-Informed PointNet framework that employs the PointNet architecture to directly process discrete sets of spatial coordinate points describing the domain boundaries. This method successfully predicts steady-state incompressible flows around multiple irregular geometries without mesh regeneration. Furthermore, implicit geometric representations, such as Signed Distance Functions (SDF), have gained widespread attention due to their continuity and resolution independence. Ghosh et al. \cite{ghosh2024geometry} extended this geometry-aware capability to turbulence prediction, proposing an embedding strategy that combines global design parameters with local SDF values, effectively achieving flow inference for unseen airfoils under turbulent conditions.

Despite significant progress in integrating geometric features with PINNs, existing methodologies remain constrained by notable limitations. Current research on PINNs is predominantly confined to canonical geometries governed by restricted parameterization, focusing largely on airfoil optimization problems defined by a limited set of shape coefficients. Furthermore, the vast majority of existing frameworks are limited to the prediction of steady-state flow fields. Consequently, there is a need to develop a more universal geometric representation strategy and integrate it with PINN architectures to enable robust prediction of unsteady flows around complex geometric structures.

In this study, we propose the Geometry-Parameterized Dual-Encoder PINN (GP-DE-PINN), a unified framework for predicting laminar flow fields around cylinders of varying geometries. The model features a dual-encoder architecture consisting of a geometric parameter encoder to extract latent geometric features and a spatiotemporal coordinate encoder to model flow dynamics. The paper is organized as follows: Section \ref{sec:method} mainly details the methodology of the GP-DE-PINN, including the dual-encoder architecture and the construction of the physics-informed loss function; Section \ref{sec:result} illustrates the prediction results, evaluating the prediction accuracy on petal-shaped cylinders; Section \ref{sec:discussion} presents the sensitivity analysis to geometric sampling density and the network width; Section \ref{sec:conclusion} concludes the study and discusses future directions.

\section{\label{sec:method}Methods}

This section details the methodological framework employed to predict laminar flow fields around cylinders of varying geometries. \Cref{sec:Geometric Parameter} describes the specific sampling strategy used to  transform continuous  boundaries into structured  geometric parameters. Subsequently, \cref{sec:Physics-Informed Neural Network} introduces the PINN that incorporates these geometric parameters as additional inputs, termed the Geometry-Parameterized PINN (GP-PINN). Furthermore, to enhance predictive accuracy, we propose the Geometry-Parameterized Dual-Encoder PINN (GP-DE-PINN), which utilizes a dual-encoder architecture to integrate spatiotemporal and geometric features, as illustrated in \cref{sec:GP-PINN}. Finally, \cref{sec:Data Set} outlines the generation of the CFD dataset, the sampling strategies for training, and the setting of the training process.

\subsection{\label{sec:Geometric Parameter}Geometric Parameter}

\begin{figure}[!htbp]
\centering
\includegraphics[width=0.5\textwidth]{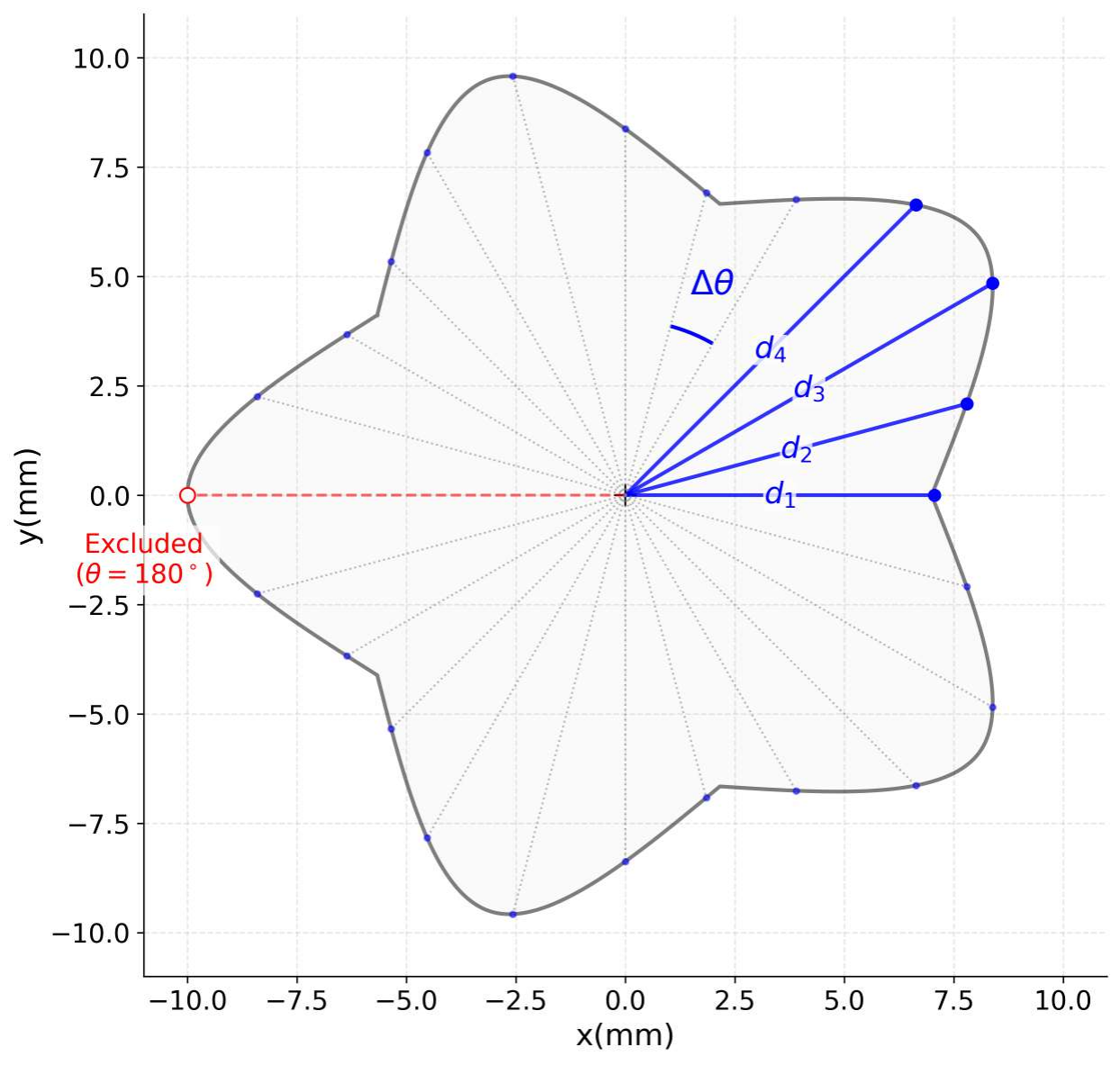}
\caption{\label{fig:geo_parameter} Schematic illustration of the geometric parameterization scheme. The continuous boundary is discretized into a high-dimensional feature vector $\mathbf{d} = [d_1, d_2, \dots, d_m]$, where each component $d_i$ represents the radial distance from the centroid to the boundary, sampled at uniformly spaced angular interval $\Delta \theta$.}
\end{figure}

To transform the continuous boundary profile into a structured numerical input compatible with the neural network, a discrete polar sampling strategy is used. As illustrated in \cref{fig:geo_parameter}, the geometry is characterized by the radial distance $d(\theta)$ from the centroid to the surface, with the azimuthal domain $\theta \in [0, 2\pi)$ discretized at a uniform angular resolution of $\Delta \theta$. During the feature extraction process, the data point corresponding to the coordinate at $\theta = 180^\circ$ is excluded. Since this coordinate serves as a fixed geometric anchor across shape variations, it contributes zero variance to the dataset. Its removal effectively reduces the input dimensionality without resulting in any loss of geometric information. Consequently, the specific geometric configuration is encapsulated in a high-dimensional feature vector defined as $\mathbf{d} = [d_1, d_2, \dots, d_{m}]$, where $m=\frac{360^{\circ}}{\Delta\theta}-1$.

\subsection{\label{sec:Physics-Informed Neural Network}GP-PINN: Geometry-Parameterized PINN}

The PINN integrates deep learning with mathematical physics by embedding governing equations directly into the network's optimization objective \cite{raissi2019physics}. Consider a general nonlinear partial differential equation defined over a spatial domain $\Omega \subset \mathbb{R}^d$ with dimension $d$ and a temporal interval $t \in [0, T]$:
\begin{equation}
u_t + \mathcal{N}[u] = 0, \quad x \in \Omega, \quad t \in [0, T],
\end{equation}
where $u(t, x)$ denotes the latent solution at time $t$ and spatial coordinate $x$. The term $u_t$ represents the partial derivative of the solution with respect to time ($\partial u / \partial t$), and $\mathcal{N}[\cdot]$ signifies a general differential operator that encompasses nonlinear spatial derivatives and physical parameters governing the system dynamics.

\begin{figure}[!htbp]
\centering
\includegraphics[width=0.65\textwidth]{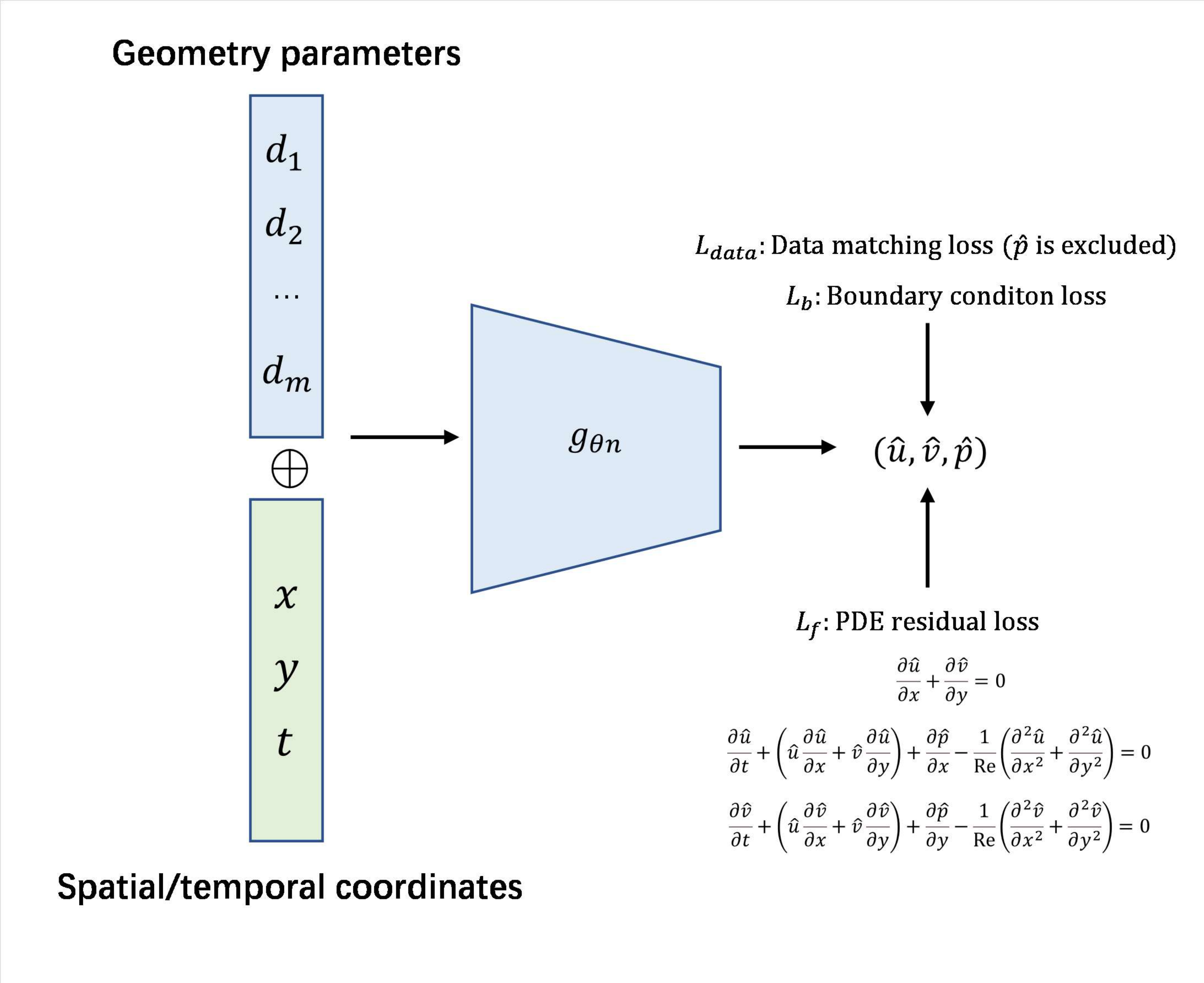}
\caption{\label{fig:pinn1} Schematic illustration of Geometry-Parameterized PINN architecture. }
\end{figure}

To approximate the exact solution $u(t, x)$, a deep neural network is employed, denoted as $\hat{u}(t, x; \phi)$, which takes the spatiotemporal coordinates $(t, x)$ as inputs and is parameterized by a set of weights and biases $\phi$. By using automatic differentiation, the derivatives of the network output with respect to the input coordinates can be precisely computed. Consequently, the physics-informed residual function $f(t, x)$ is defined as:
\begin{equation}
f(t, x) := \frac{\partial \hat{u}}{\partial t} + \mathcal{N}[\hat{u}].
\end{equation}
This residual quantifies the discrepancy between the neural network's prediction and the governing physical laws. The network parameters $\phi$ are optimized by minimizing a composite loss function $\mathcal{L}(\phi)$, which enforces the PDE constraints, fits the observed measurement data, and satisfies the boundary conditions:
\begin{equation}
\mathcal{L}(\phi) = w_f \mathcal{L}f + w_{data} \mathcal{L}_{data} + w_b \mathcal{L}_b.
\end{equation}
In this formulation, $\mathcal{L}_f$, $\mathcal{L}_{data}$, and $\mathcal{L}_b$ represent the loss terms associated with the PDE residual, the labeled data, and the boundary conditions, respectively. The non-negative coefficients $w_f$, $w_{data}$, and $w_b$ are weights used to balance the relative contribution of each term during the training process. These loss components are typically formulated as Mean Squared Errors (MSE) over discrete sets of training points:
\begin{equation}
\mathcal{L}_f = \frac{1}{N_f} \sum_{i=1}^{N_f} \left| f(t_f^i, x_f^i) \right|^2,
\end{equation}
\begin{equation}
\mathcal{L}_{data} = \frac{1}{N_{data}} \sum_{i=1}^{N_{data}} \left| \hat{u}(t_{data}^i, x_{data}^i) - u_{data}^i \right|^2,
\end{equation}
\begin{equation}
\mathcal{L}_b = \frac{1}{N_b} \sum_{i=1}^{N_b} \left| \hat{u}(t_b^i, x_b^i) - g(t_b^i, x_b^i) \right|^2.
\end{equation}
Here, $N_f$ denotes the number of collocation points $(t_f^i, x_f^i)$ sampled within the domain $\Omega$ to enforce the PDE structure. $N_{data}$ represents the number of observed data points $(t_{data}^i, x_{data}^i)$ where the ground truth solution values $u_{data}^i$ are known. Similarly, $N_b$ is the number of points $(t_b^i, x_b^i)$ sampled on the domain boundary $\partial \Omega$, where the predicted solution should match the prescribed boundary function $g(t, x)$. By minimizing $\mathcal{L}(\phi)$, the network converges to a solution that simultaneously satisfies the governing physical equations and complies with the available data and the boundary constraints.

To enable PINN to predict the flow field around obstacles of different shapes, we first attempt to use geometric parameters as additional inputs for PINN. As shown in the \cref{fig:pinn1}. Geometry-Parameterized PINN (GP-PINN) concatenates the geometric parameter vector $\mathbf{d}$ with the spatiotemporal coordinates $(x,y,t)$. This unified vector is then propagated through a fully connected neural network $g_{\theta n}$ to approximate the flow variables $(\hat{u}, \hat{v}, \hat{p})$. The network parameters are optimized by minimizing a composite loss function, which enforces data matching loss $L_{data}$, boundary condition loss $L_{b}$, and the physical constraints imposed by the Navier-Stokes PDE residuals $L_{f}$.

\subsection{\label{sec:GP-PINN}GP-DE-PINN: Geometry-Parameterized Dual-Encoder PINN}


To enhance prediction accuracy, we propose the Geometry-Parameterized Dual-Encoder PINN, an extension of the GP-PINN framework. The distinction of GP-DE-PINN lies in its dual-encoder architecture. The detailed structure of this model is introduced as follows.

\subsubsection{\label{sec:level3}Network Overview}

\begin{figure}[!htbp]
\centering
\includegraphics[width=1\textwidth]{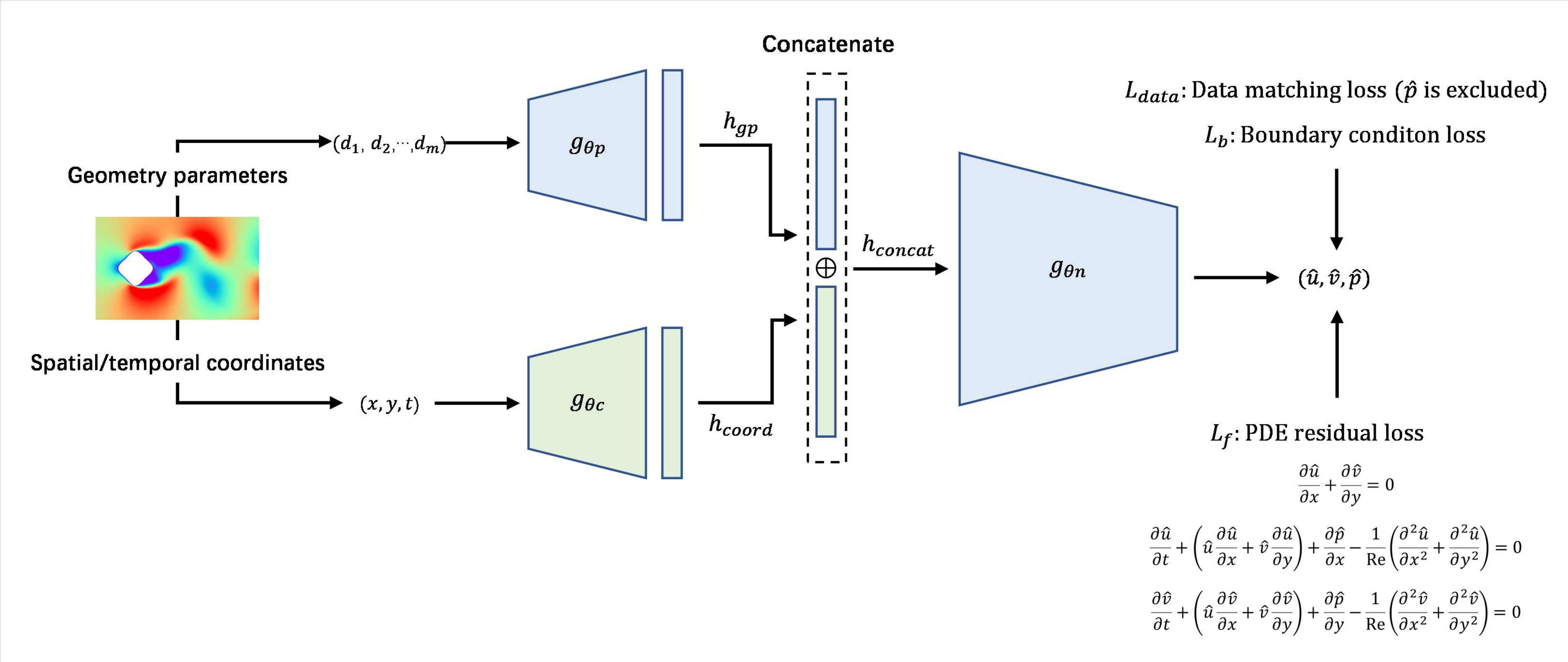}
\caption{\label{fig:gp-pinn_pic} GP-DE-PINN architecture. The two encoders $g_{\theta p}$ and $g_{\theta c}$ are employed to generate enhanced representations for the geometric parameters and the spatial/temporal coordinates. The concatenated features are then processed by the prediction network $g_{\theta n}$. The Navier-Stokes equations are provided as the governing physical constraints.}
\end{figure}

As illustrated in \cref{fig:gp-pinn_pic}, the network comprises three distinct functional modules: a geometric parameter encoder $g_{\theta_{p}}$, a spatiotemporal coordinate encoder $g_{\theta_{c}}$, and a manifold approximation network $g_{\theta_{n}}$ \cite{cho2024parameterized}. The overall forward propagation can be formalized as a composite mapping:
\begin{equation}
[\hat{u}, \hat{v}, \hat{p}]^T = g_{\theta_n} \left( \mathbf{h}_{coord} \oplus \mathbf{h}_{gp} \right),
\end{equation}
where $\mathbf{h}_{coord} = g_{\theta_c}(\mathbf{x})$ and $\mathbf{h}_{gp} = g_{\theta_p}(\mathbf{d})$ represent the latent feature vectors extracted from the spatiotemporal coordinates $\mathbf{x} = (x, y, t)$ and the geometric parameter vector $\mathbf{d}$, respectively. The symbol $\oplus$ denotes the vector concatenation operation, and $\Phi = \{\phi_{c}, \phi_{p}, \phi_{n}\}$ encompasses all learnable weight matrices and bias vectors within the sub-networks.

\subsubsection{\label{sec:level3}Geometric parameter Encoder}

The geometry encoder $g_{\theta_{p}}$ within the GP-DE-PINN framework is designed to map the information of the petal‑shaped cylinder boundary into a latent embedding. It takes the geometry vector $\mathbf{d}$ as input. This sub-network is constructed as a fully connected Multi-Layer Perceptron (MLP) with a depth of $K_p$. Let $\mathbf{h}_p^{(l)}$ denote the output of the $l$-th layer. The forward pass is defined recursively as:
\begin{equation}
\begin{cases}
\mathbf{h}_p^{(0)} = \mathbf{d}, \\
\mathbf{h}_p^{(l)} = \sigma \left( \mathbf{W}_p^{(l)} \mathbf{h}_p^{(l-1)} + \mathbf{b}_p^{(l)} \right), \quad \text{for } l = 1, \dots, K_p,
\end{cases}
\end{equation}
where $\mathbf{W}_p^{(l)}$ and $\mathbf{b}_p^{(l)}$ are the weight matrix and bias vector for the $l$-th layer, and $\sigma(\cdot)$ represents the activation function. The final output $\mathbf{h}_{gp} = \mathbf{h}_p^{(K_p)}$ serves as the geometric embedding.

\subsubsection{\label{sec:level3}Spatiotemporal Coordinate Encoder}

Parallel to the geometry branch, the coordinate encoder $g_{\theta_{c}}$ is responsible for processing spatiotemporal information. It transforms the physical coordinate vector $\mathbf{x} = (x, y, t)$ into a high-dimensional feature space. This sub-net comprises $K_c$ fully connected layers. Let $\mathbf{h}_c^{(l)}$ denote the output of the $l$-th layer in this branch. The encoding process is formulated recursively as follows:
\begin{equation}
\begin{cases}
\mathbf{h}_c^{(0)} = \mathbf{x}, \\
\mathbf{h}_c^{(l)} = \sigma \left( \mathbf{W}_c^{(l)} \mathbf{h}_c^{(l-1)} + \mathbf{b}_c^{(l)} \right), \quad \text{for } l = 1, \dots, K_c,
\end{cases}
\end{equation}
where $\mathbf{W}_c^{(l)}$ and $\mathbf{b}_c^{(l)}$ represent the weight matrix and the bias vector for the $l$-th layer, respectively. The final output $\mathbf{h}_{coord} = \mathbf{h}_c^{(K_c)}$ serves as the spatiotemporal feature embedding.

\subsubsection{\label{sec:level3}Manifold Network and Prediction}

The core inference is executed by the manifold network $g_{\theta_{n}}$, which functions as a fluid dynamics decoder. The latent geometric embedding $\mathbf{h}_{gp}$ and the spatiotemporal features $\mathbf{h}_{coord}$ are first fused via concatenation to form a unified state vector $\mathbf{h}_{concat}$:
\begin{equation}
\mathbf{h}_{concat} = \text{concat}(\mathbf{h}_{coord}, \mathbf{h}_{gp}) \in \mathbb{R}^{d_{total}}.
\end{equation}
This fused vector is then propagated through a deep neural network with $K_n$ layers. Let $\mathbf{h}_n^{(l)}$ be the output of the $l$-th layer. The prediction process is defined as:
\begin{equation}
\begin{cases}
\mathbf{h}_n^{(0)} = \mathbf{h}_{concat}, \\
\mathbf{h}_n^{(l)} = \sigma \left( \mathbf{W}_n^{(l)} \mathbf{h}_n^{(l-1)} + \mathbf{b}_n^{(l)} \right), \quad \text{for } l = 1, \dots, K_n-1, \\
(\hat{u}, \hat{v}, \hat{p}) = \mathbf{W}_n^{(K_n)} \mathbf{h}_n^{(K_n-1)} + \mathbf{b}_n^{(K_n)}.
\end{cases}
\end{equation}
The network is trained by minimizing a composite loss function $\mathcal{L}_{total}$, which integrates data-driven errors ($\mathcal{L}_{data}$), boundary‑condition penalties ($\mathcal{L}_{b}$), and physics-informed residuals ($\mathcal{L}_{f}$). Specifically, the term $\mathcal{L}_{f}$ is used to calculate the residuals of the equations for the incompressible flow, defined by the dimensionless continuity equation and the momentum equations for the $u$ and $v$ components:
\begin{equation}
f_{c} = \frac{\partial \hat{u}}{\partial x} + \frac{\partial \hat{v}}{\partial y} = 0,
\end{equation}
\begin{equation}
f_{u} = \frac{\partial \hat{u}}{\partial t} + \hat{u} \frac{\partial \hat{u}}{\partial x} + \hat{v} \frac{\partial \hat{u}}{\partial y} + \frac{\partial \hat{p}}{\partial x} - \frac{1}{Re} \left( \frac{\partial^2 \hat{u}}{\partial x^2} + \frac{\partial^2 \hat{u}}{\partial y^2} \right) = 0,
\end{equation}
\begin{equation}
f_{v} = \frac{\partial \hat{v}}{\partial t} + \hat{u} \frac{\partial \hat{v}}{\partial x} + \hat{v} \frac{\partial \hat{v}}{\partial y} + \frac{\partial \hat{p}}{\partial y} - \frac{1}{Re} \left( \frac{\partial^2 \hat{v}}{\partial x^2} + \frac{\partial^2 \hat{v}}{\partial y^2} \right) = 0,
\end{equation}
where $Re$ represents the Reynolds number. All partial derivatives in these residuals are computed by automatic differentiation \cite{baydin2018automatic}. It should be noted that, only the velocity field data (the $u$ and $v$ components) are used during training, while the pressure field data ($p$) are excluded. Thus, the data loss term $\mathcal{L}_{data}$ and boundary loss term $\mathcal{L}_{b}$ are calculated solely based on the velocity components. Consequently, the total loss function is formulated as follows:
\begin{equation}
\mathcal{L}_{total} = \lambda_{data}\mathcal{L}_{data}(\hat{u}, \hat{v}) + \lambda_{b}\mathcal{L}_{b}(\hat{u}, \hat{v}) + \lambda_{f}\mathcal{L}_{f}(\hat{u}, \hat{v},\hat{p}).
\end{equation}
The specific hyperparameter configuration employed in this work is as follows: the geometric parameter encoder $g_{\theta_{p}}$ consists of $K_p = 4$ layers. The spatiotemporal coordinate encoder $g_{\theta_{c}}$ is composed of $K_c = 3$ layers. The manifold network $g_{\theta_{n}}$ consists of $K_n = 5$ layers. The numbers $N_p$, $N_c$, and $N_n$ of hidden layer neurons in $g_{\theta_{p}}$, $g_{\theta_{c}}$, and $g_{\theta_{n}}$ are 250, 50 and 100, respectively. The hyperbolic tangent function is chosen as the activation function. Furthermore, the weighting coefficients for the three loss components are set to $\lambda_{data} = 1.0$, $\lambda_{b} = 1.0$, and $\lambda_{f} = 1.0$.

\subsection{\label{sec:Data Set}Data Set}

\begin{figure}[!h]
\centering
\includegraphics[width=0.5\textwidth]{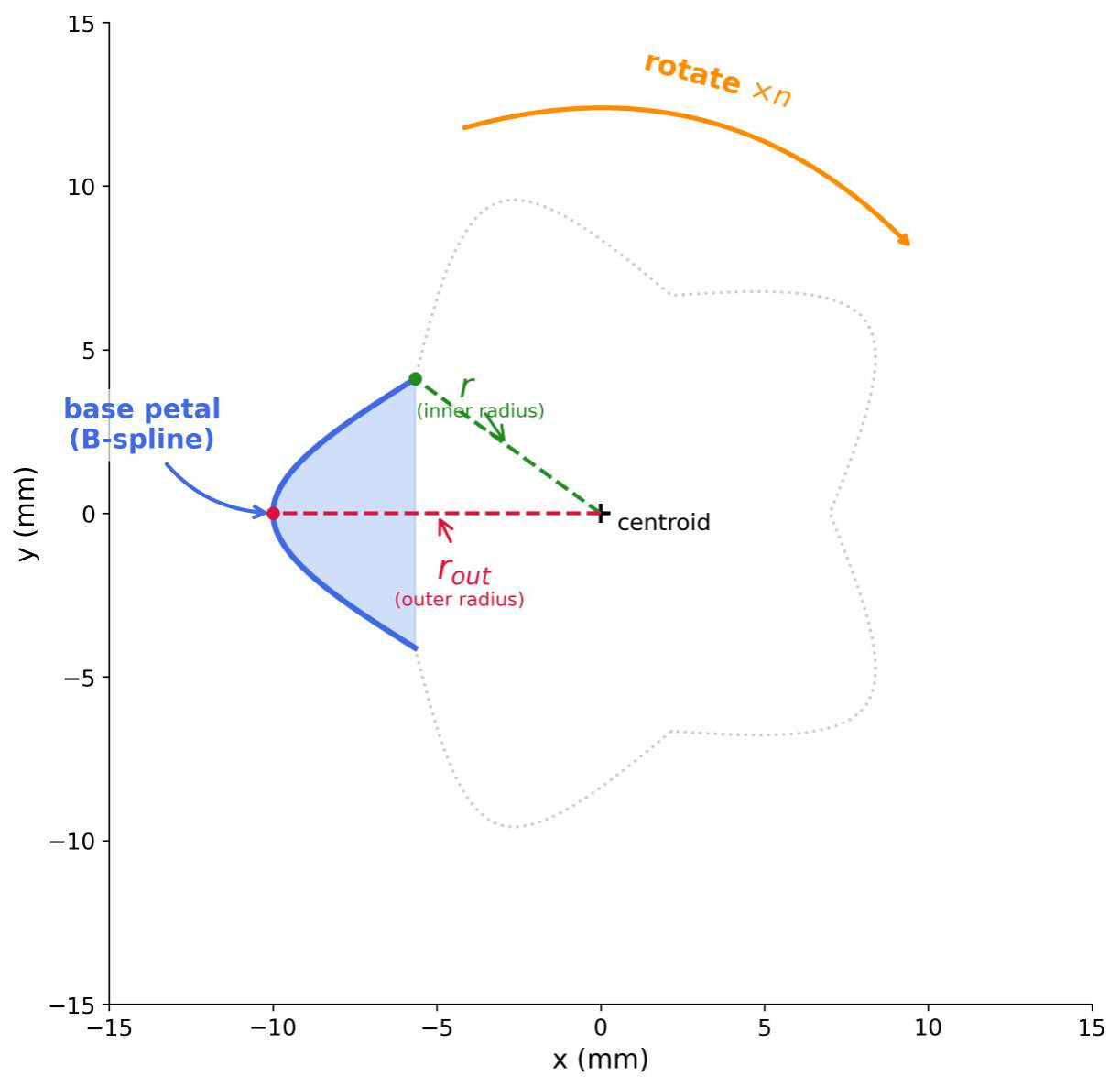}
\caption{\label{fig:Data Set} Methods for generating petal-shaped cylinders.}
\end{figure}

To evaluate the model's generalization capability, a dataset of 45 distinct 2D petal-shaped cylinders is constructed using a parametric rotational assembly method. As illustrated in \cref{fig:Data Set}, the boundary variation is governed by two key variables: the inner radius ($r$) and the number of petals ($n$). For each configuration, a fundamental "base petal" is defined by vertices that change between a variable inner radius $r$ and a fixed outer radius ($r_{out}=10.0$ mm). To ensure the smoothness and continuity of the boundaries, B-spline curves are employed to generate a smooth boundary profile based on these vertices. The complete closed-loop boundary is then generated by rotating this smooth base profile $n$ times around the origin. The data set covers a geometric space by varying the inner radius $r \in [7.0, 9.0]$ mm with a step size of 0.25, and the petal count $n \in \{4, 5, 6, 7, 8\}$, thus creating a spectrum of shapes.

\begin{figure}[!h]
\centering
\includegraphics[width=1\textwidth]{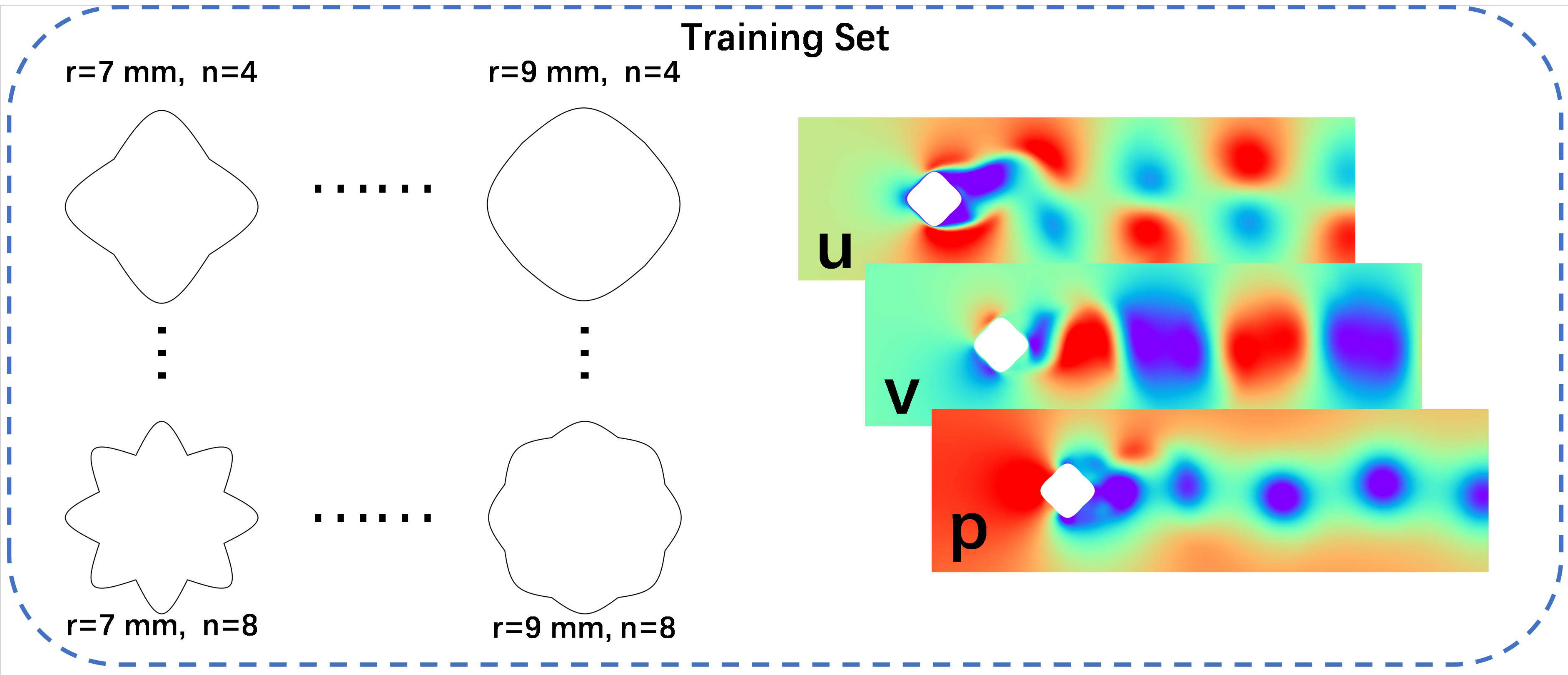}
\caption{\label{fig:Training Set} Training sets for evaluating the GP-PINN and the GP-DE-PINN.}
\end{figure}

\begin{figure}[!h]
\centering
\includegraphics[width=1\textwidth]{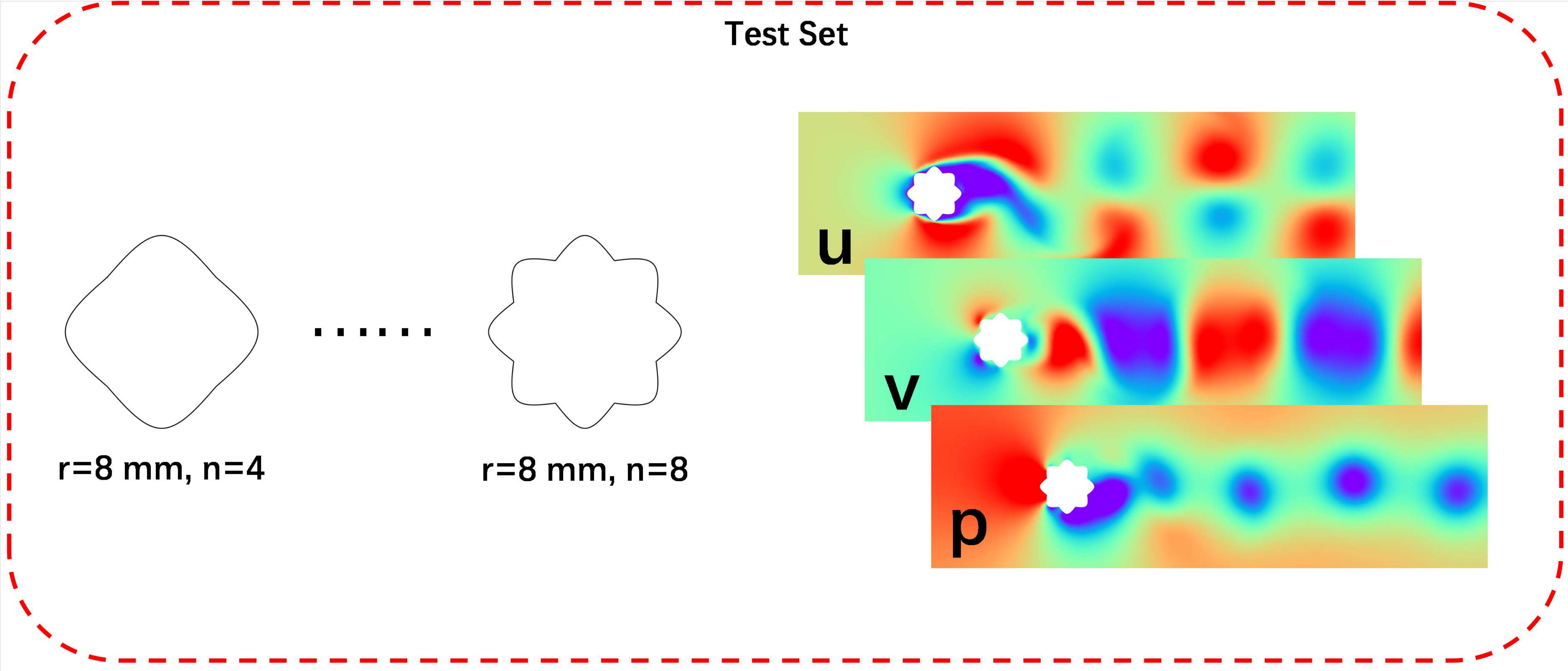}
\caption{\label{fig:Test Set} Test sets for evaluating the GP-PINN and the GP-DE-PINN.}
\end{figure}

From this complete data set of 45 cylinders, the five configurations corresponding to an inner radius of $r=8.0$ mm with varying petal counts are selected as the test set to assess model performance in unseen geometries. The remaining 40 geometries are utilized for training, as illustrated in \cref{fig:Training Set} and \cref{fig:Test Set}. The GP-DE-PINN and GP-PINN are evaluated using flow-field data with consistent physical properties: a fluid density of $\rho = 1000$ kg/m$^3$ and a free-stream velocity of $U_\infty = 0.1$ m/s. The characteristic length for all corrugated cylinder configurations is fixed at $D = 0.02$ m. Correspondingly, the Reynolds number is maintained at $Re = 400$. Our study restricts the flow to this specific laminar condition to isolate and clearly evaluate the impact of geometric variations on flow patterns. 

We solve the incompressible Navier-Stokes equations using STAR CCM+ to obtain high-fidelity ground truth flow fields around the 2D petal-shaped cylinders. The analysis is confined to a computational window defined by $[x_{min}, x_{max}] \times [y_{min}, y_{max}] = [-0.05, 0.155] \times [-0.03, 0.03]$ m. Each simulation spans a physical duration of 8 s, capturing the velocity components ($u, v$) and pressure ($p$).

For model training, we employ a stratified randomized sampling strategy. Specifically, 2,000 spatiotemporal collocation points are sampled from each of the 40 training cases, culminating in a global interior data set of 80,000 points. Additionally, to accurately compute the loss of boundary condition ($L_{b}$), we construct a specialized boundary dataset by extracting 5,000 points from the cylinder walls, 5,000 points from the inlet boundary, and 5,000 points from the flow field at the initial timestamp in all training instances, generating a total of 15,000 boundary points.

The network parameters $\Theta$ are optimized to minimize the composite loss function comprising data matching errors, the constraints of boundary conditions, and physics-informed residuals.We employ the L-BFGS optimizer \cite{liu1989limited}, a second-order quasi-Newton method, to accelerate convergence and achieve high-precision solutions. Optimization is conducted for 50,000 iterations using a full-batch training strategy. The network weights are initialized using the Xavier normal method \cite{glorot2010understanding}, while the biases are initialized to zero.



\section{\label{sec:result}Results}

We evaluate the GP-DE-PINN model in training and test sets that contain obstacles. Due to the interference of small perturbations in the two-dimensional cylinder flow, there is a phase difference between the flow field predicted by the model and the true flow field \cite{hu2025generative}. Therefore, when the flow becomes stable, a set of snapshots is selected to represent the complete period within the simulation time and compare them with the model output. Similar to \citet{hu2025generative}, we use the $L_1$ error to measure the similarity between the model outputs and the snapshots in the ground truth.
\begin{equation}L_{1}(V_{0}^{M})=\|V_{0}^{M}-V_{0}^{\mathrm{GT}}\|,\end{equation}

Here, $V$ represents the velocity magnitude, and the superscript \(M\) denotes the model output. For each test set, the snapshot of the predicted flow field that has the minimum \(L_{1}(V_{0}^{M})\) with respect to the first snapshot of complete period in ground truth, is used for subsequent comparison over the entire period.

\subsection{\label{sec:level2}Training Set}

\begin{figure}[!h]
  \centering
  \subfloat[]{\includegraphics[width=1\textwidth]{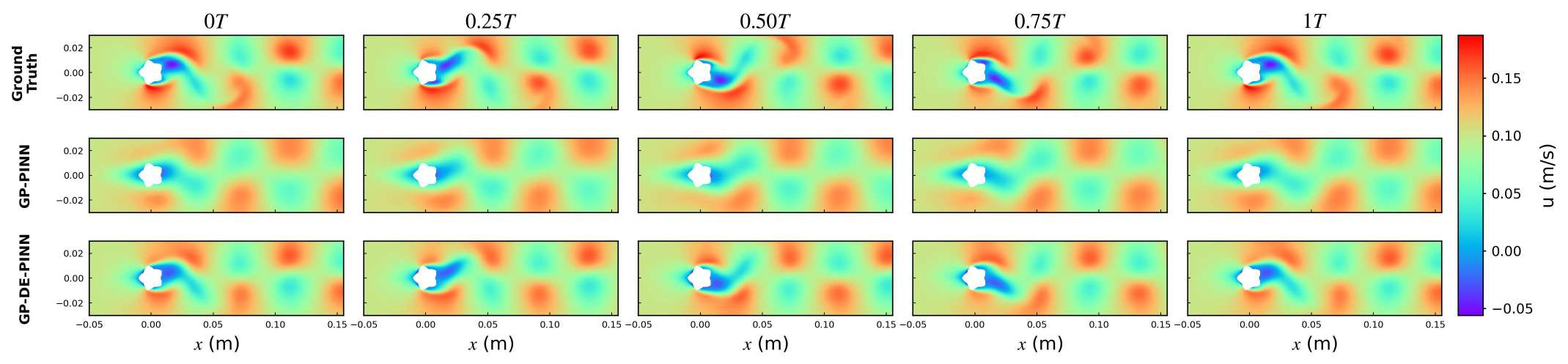}} \\
  \subfloat[]{\includegraphics[width=1\textwidth]{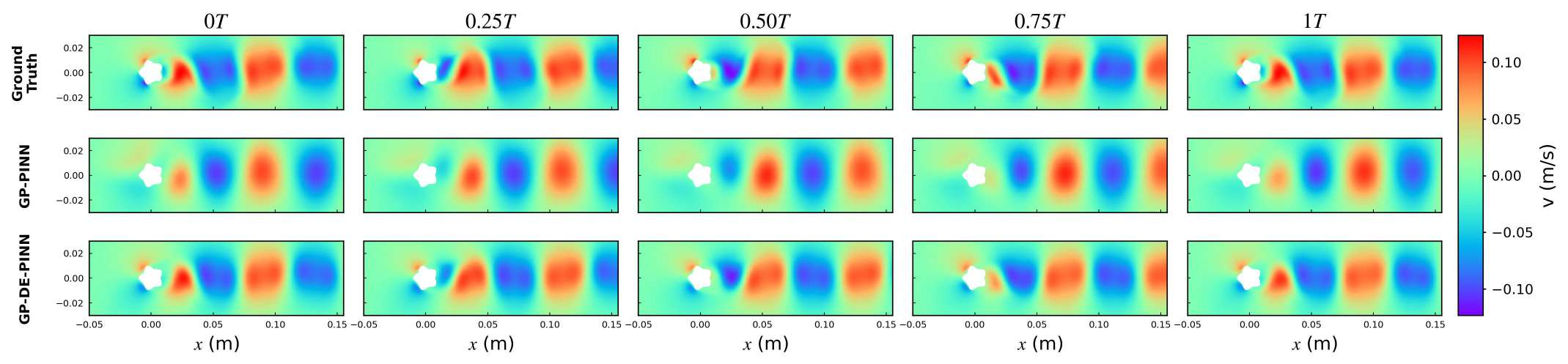}} \\
  \subfloat[\label{7.5-5pressure}]{\includegraphics[width=1\textwidth]{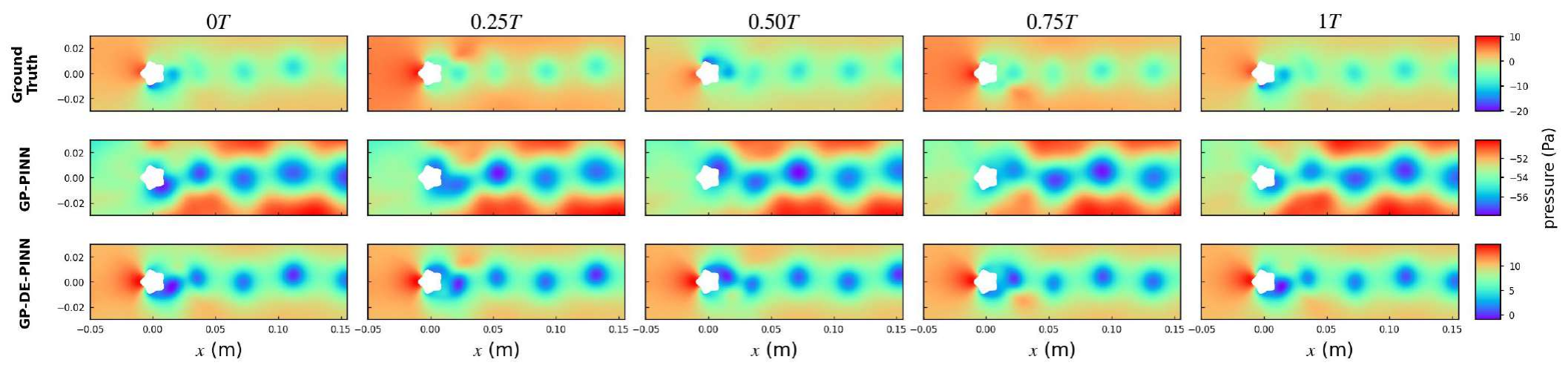}} 
  \caption{\label{7.5-5pinn&gppinn} The flow fields around the petal-shaped cylinder ($r=7.5$ mm, $n=5$) in the training sets generated by the GP-PINN and GP-DE-PINN over the whole period, compared with the ground truth. (a) Contour of $u$. (b) Contour of $v$. (c) Contour of $p$. }
\end{figure}

\begin{figure}[!h]
  \centering
  \subfloat[]{\includegraphics[width=1\textwidth]{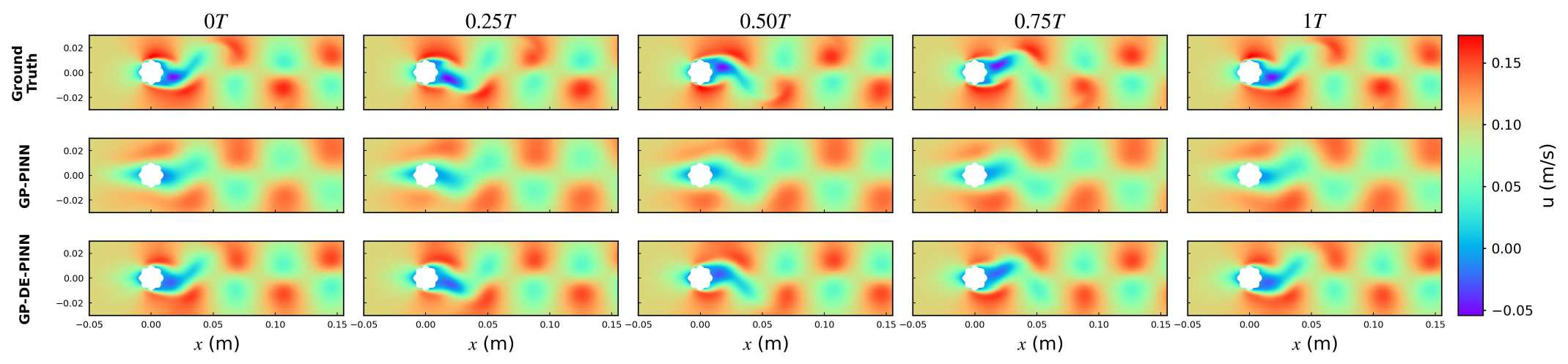}} \\
  \subfloat[]{\includegraphics[width=1\textwidth]{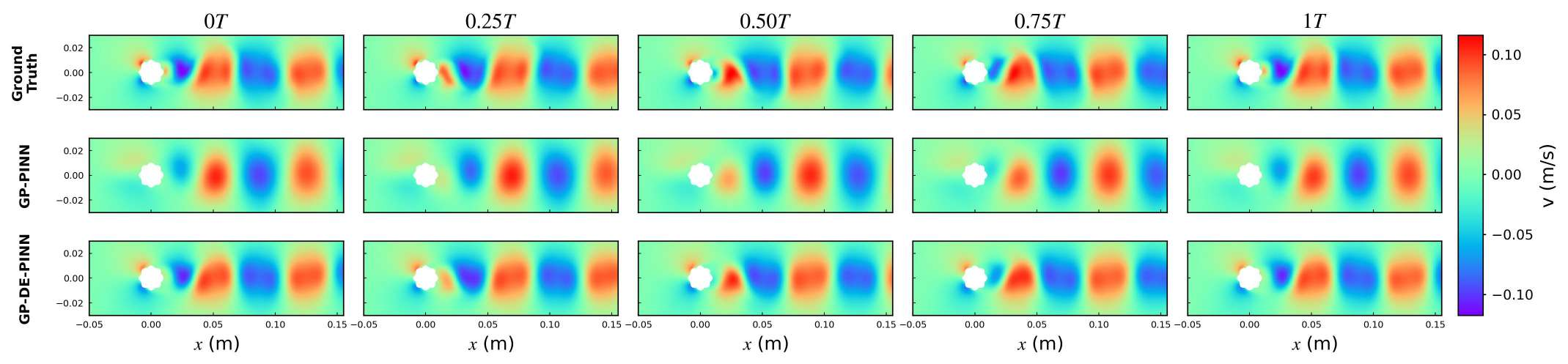}} \\
  \subfloat[\label{8.5-8pressure}]{\includegraphics[width=1\textwidth]{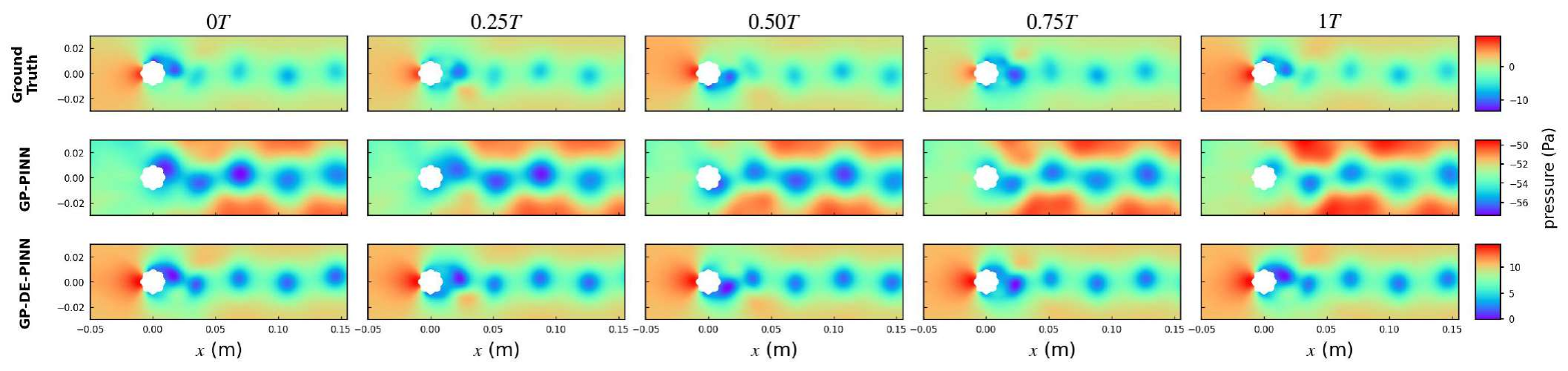}}
  \caption{\label{8.5-8pinn&gppinn} The flow fields around the petal-shaped cylinder ($r=8.5$ mm, $n=8$) in the training sets generated by the GP-PINN and GP-DE-PINN over the whole period, compared with the ground truth. (a) Contour of $u$. (b) Contour of $v$. (c) Contour of $p$. }
\end{figure}

The reconstruction capability of the models is evaluated using samples from the training set, taking two cases with distinct geometric parameters as examples: case A ($r=7.5$ mm, $n=5$) and case B ($r=8.5$ mm, $n=8$). \Cref{7.5-5pinn&gppinn} and \cref{8.5-8pinn&gppinn} present the spatiotemporal evolution of the velocity components ($u, v$) and the pressure field ($p$) over a complete vortex shedding cycle ($T$). The comparison includes the ground truth, the GP-PINN prediction and the GP-DE-PINN prediction. Regarding the velocity fields, while the GP-PINN manages to capture the fundamental periodicity of the Karman vortex street, it exhibits noticeable numerical dissipation. As the vortices convect downstream, the GP-PINN model produces "blurred" flow structures with attenuated velocity magnitudes. In contrast, the GP-DE-PINN demonstrates superior fidelity. It accurately reconstructs the details of the flow field and maintains the sharpness of the shear layers.

In addition, the model's capability to predict the pressure field is investigated. It is noteworthy that the pressure data is not used during training, while the model can still provide the pressure field prediction through the constraints of the physical governing equations. The training losses ($\mathcal{L}_{data}$ and $\mathcal{L}_{b}$) exclude pressure data, relying solely on velocity measurements. Consequently, as shown in \cref{7.5-5pressure} and \cref{8.5-8pressure}, certain numerical discrepancies in the absolute magnitude of the predicted pressure are observed when compared to the ground truth. However, despite the absence of direct pressure supervision, the GP-DE-PINN successfully captures the correct trends of spatiotemporal variation and the structural evolution of the pressure field. This capability confirms that the network effectively infers the latent pressure variable by strictly adhering to the momentum conservation constraints embedded within the Navier-Stokes equations.

A quantitative analysis is conducted on two representative geometric cases from the training set. As shown in \cref{tab:training_comparison}, the GP-PINN exhibits relatively high residual levels when handling these complex geometric configurations, with root mean square error (RMSE) values for both $u$ and $v$ components generally hovering between $0.016$ and $0.017$, and the $v$-component mean relative error (MRE) consistently exceeding $24\%$. In contrast, the GP-DE-PINN achieves significant reductions in both absolute and relative errors. Taking the $n=5$ case as an instance, the GP-DE-PINN decreases the $u$-component RMSE from $0.017$ to $0.009$, and effectively halves the $v$-component RMSE from $0.017$ to $0.008$. This improvement directly translates to a drastic cut in the $v$-component MRE from $26.10\%$ to $12.33\%$. The performance gain is even more evident in the $n=8$ case, where the RMSE for $u$ and $v$ components drops further to $0.007$ and $0.006$, respectively, successfully suppressing the $v$-component relative error to $9.31\%$, far outperforming the GP-PINN's $24.58\%$.

\begin{table}[!h]
\caption{\label{tab:training_comparison} Comparison of RMSE and MRE for $u$ and $v$ velocity components in the training set cases.}
\begin{ruledtabular}
\begin{tabular}{cccccc}
 & & \multicolumn{2}{c}{$u$-component} & \multicolumn{2}{c}{$v$-component} \\
 \cline{3-4} \cline{5-6}
 Case & Model & RMSE & MRE (\%) & RMSE & MRE (\%) \\
 \hline
\multirow{2}{*}{\makecell{$r=7.5$mm \\ $n=5$}}
    & GP-PINN    & 0.017 & 12.01 & 0.017 & 26.10 \\
    & GP-DE-PINN & \textbf{0.009} & \textbf{6.20}  & \textbf{0.008} & \textbf{12.33} \\
 \hline
 \multirow{2}{*}{\makecell{$r=8.5$mm \\ $n=8$}}
    & GP-PINN    & 0.016 & 11.41 & 0.015 & 24.58 \\
    & GP-DE-PINN & \textbf{0.007} & \textbf{4.53}  & \textbf{0.006} & \textbf{9.31} \\
\end{tabular}
\end{ruledtabular}
\end{table}

\begin{figure}[!h]
  \centering
  \subfloat[]{\includegraphics[width=0.495\textwidth]{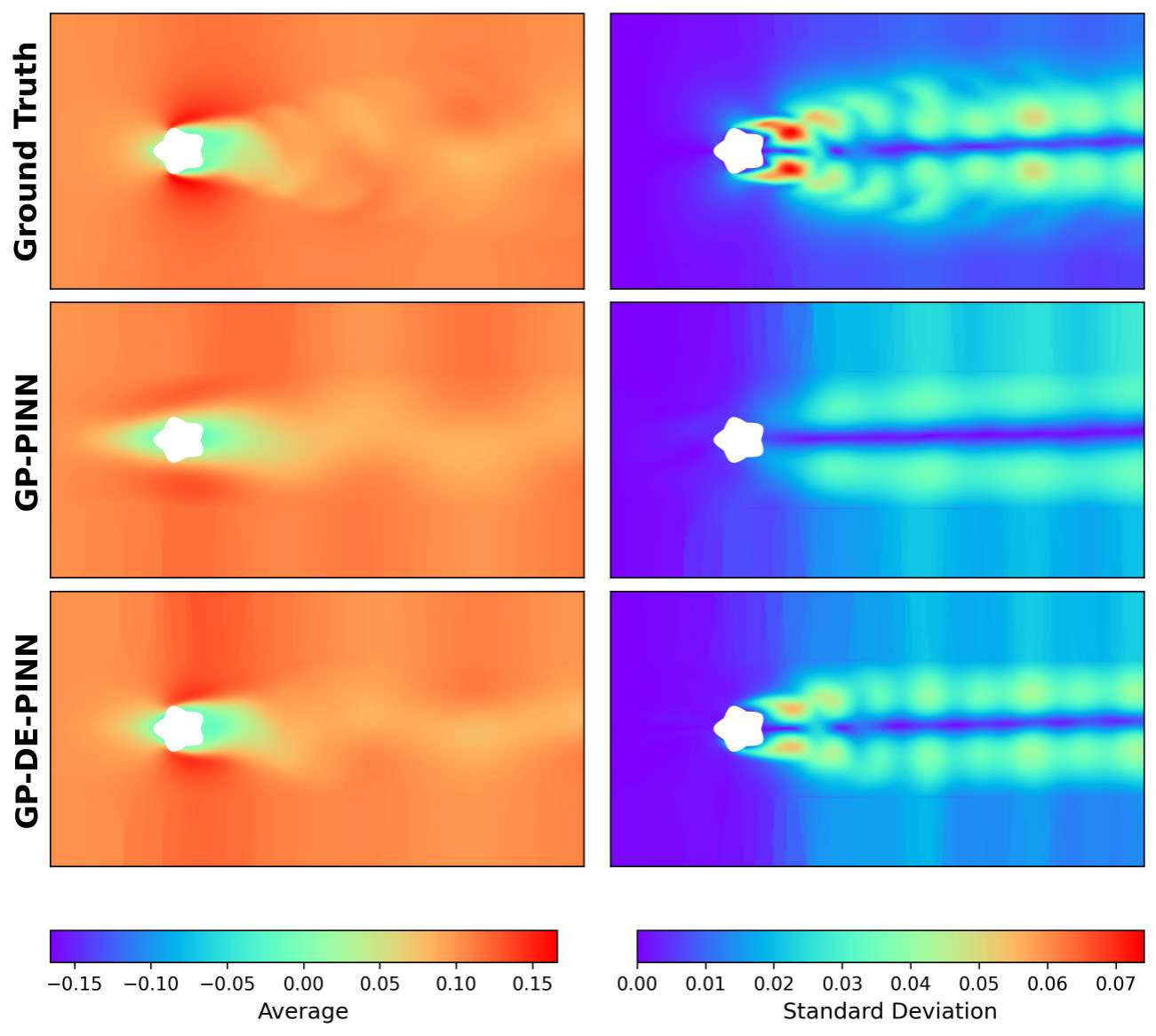}} 
  \subfloat[]{\includegraphics[width=0.495\textwidth]{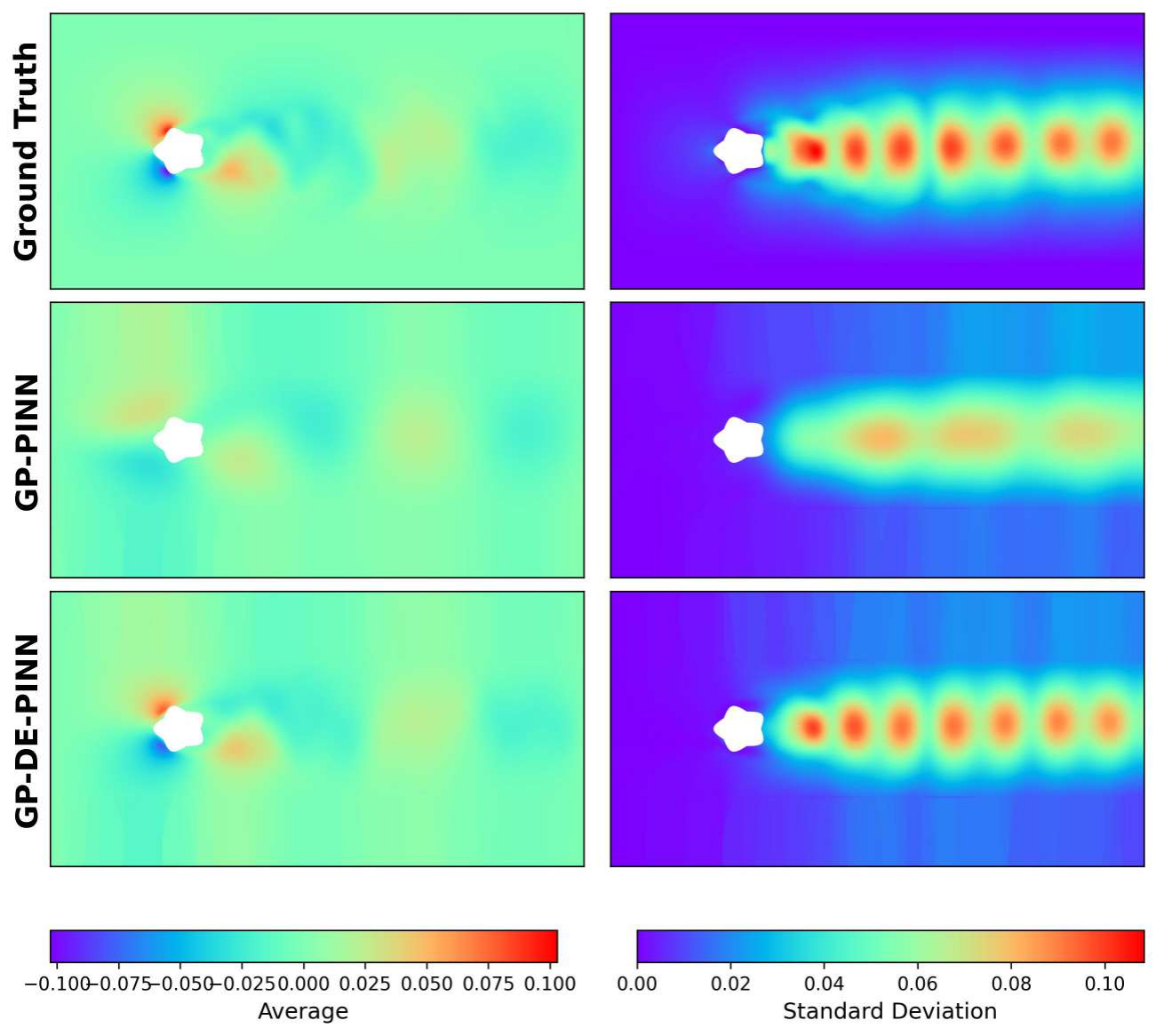}}
  \caption{\label{7.5-5avg&std} Statistical evaluation of GP-PINN and GP-DE-PINN on flow around the petal-shaped cylinder ($r=7.5$ mm, $n=5$) in the training sets. Statistics are computed over the whole period.  (a) Mean and standard deviation of $v$. (b) Mean and standard deviation of $u$. }
\end{figure}

\begin{figure}[!h]
  \centering
  \subfloat[]{\includegraphics[width=0.495\textwidth]{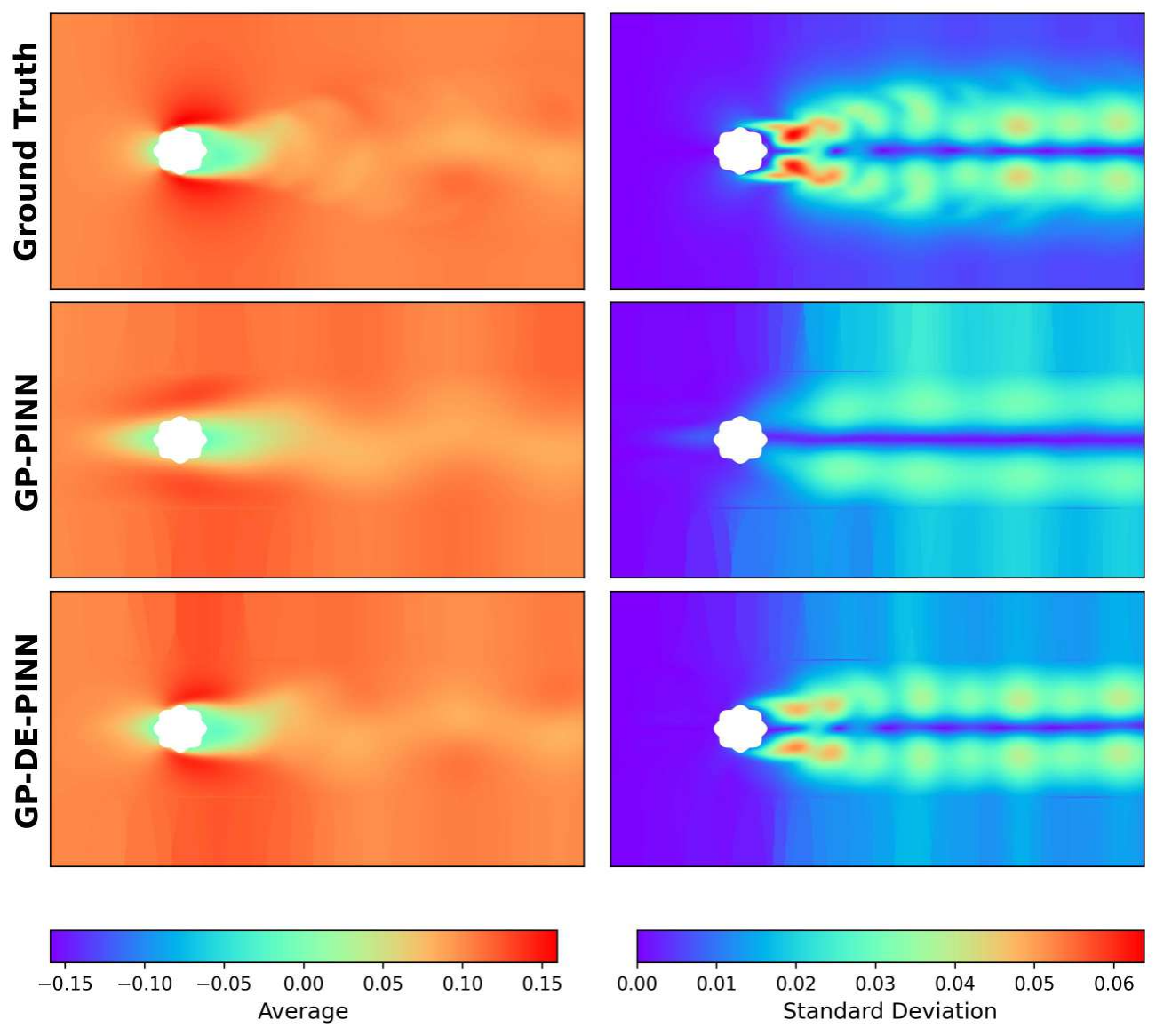}} 
  \subfloat[]{\includegraphics[width=0.495\textwidth]{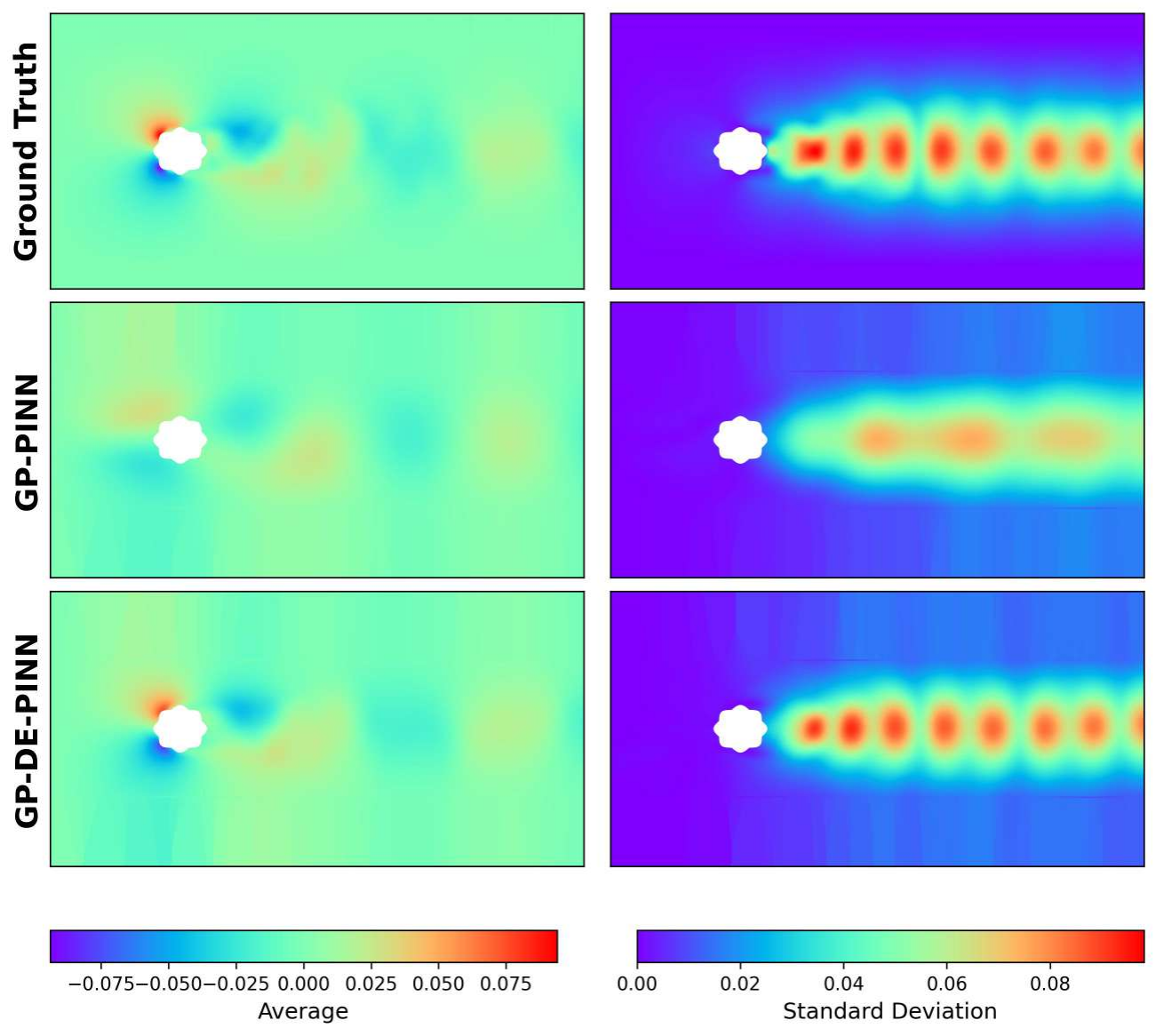}}
  \caption{\label{8.5-8avg&std} Statistical evaluation of GP-PINN and GP-DE-PINN on flow around the petal-shaped cylinder ($r=8.5$ mm, $n=8$) in the training sets. Statistics are computed over the whole period.  (a) Mean and standard deviation of $u$. (b) Mean and standard deviation of $v$. }
\end{figure}

To further quantify the performance of the model, the time-averaged velocity fields and their corresponding standard deviations are analyzed, as presented in \cref{7.5-5avg&std} and \cref{8.5-8avg&std}. In terms of mean flow ($\bar{u}, \bar{v}$), compared to GP-PINN, GP-DE-PINN achieves a more accurate reconstruction of the flow field. Moreover, a critical disparity arises in the prediction of second-order statistics, which represent the fluctuation intensity of the flow. The standard deviation fields of ground truth exhibit distinct, bead-like high-variance zone, corresponding to the shedding vortices. The GP-PINN fails to resolve these high-frequency features, resulting in a smeared, continuous band of variance that underestimates the true intensity of laminar flow. In contrast, GP-DE-PINN successfully reproduces the discrete lobe structures and the similar intensity decay rates in the far wake. These results confirm that the GP-DE-PINN not only captures the mean flow but also the complex fluctuations, making it a reliable surrogate model for reconstructing complex unsteady flows.

\subsection{\label{sec:level2}Test Set}

Following the validation of the model's reconstruction accuracy on the training dataset, we proceed to evaluate the generalization performance of the GP-DE-PINN framework. A robust surrogate model should possess the capability to accurately predict flow fields for geometric configurations that are explicitly excluded from the training phase.

To evaluate generalization, a test set consisting of five different cylinder configuration is employed. These geometries share a fixed inner radius of $r = 8.0$ mm, with the number of petals varying throughout $n \in \{4, 5, 6, 7, 8\}$. It is noted that these five cases are excluded during training process. This setup serves to validate the generalization capability of the GP-DE-PINN framework on unseen geometric configurations.

\begin{figure}[!htb]
  \centering
  \subfloat[]{\includegraphics[width=1\textwidth]{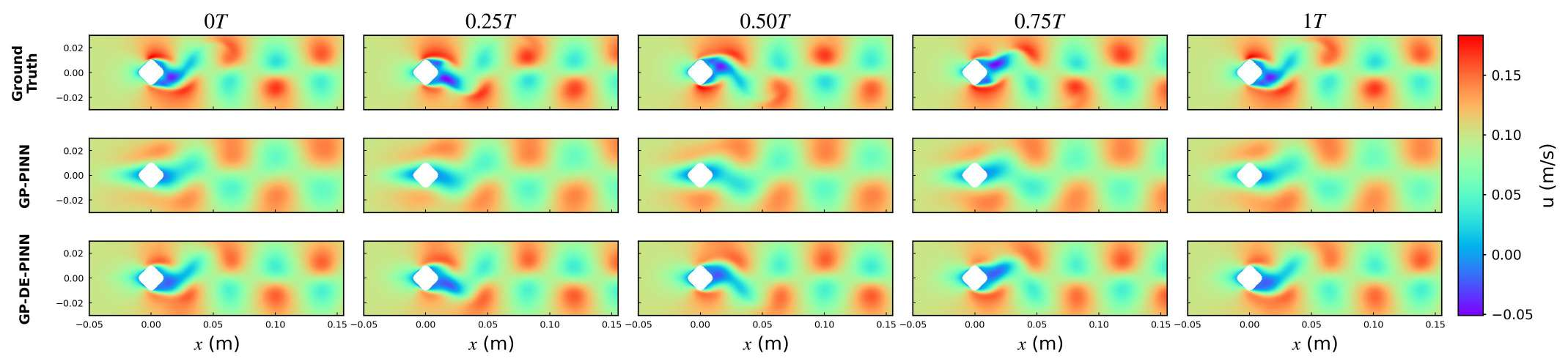}} \\
  \subfloat[]{\includegraphics[width=1\textwidth]{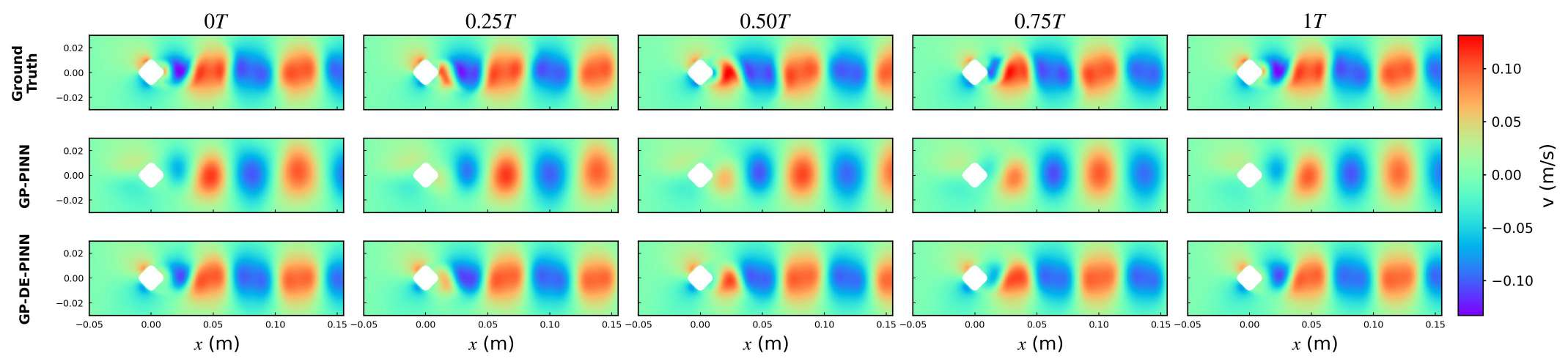}} \\
  \subfloat[]{\includegraphics[width=1\textwidth]{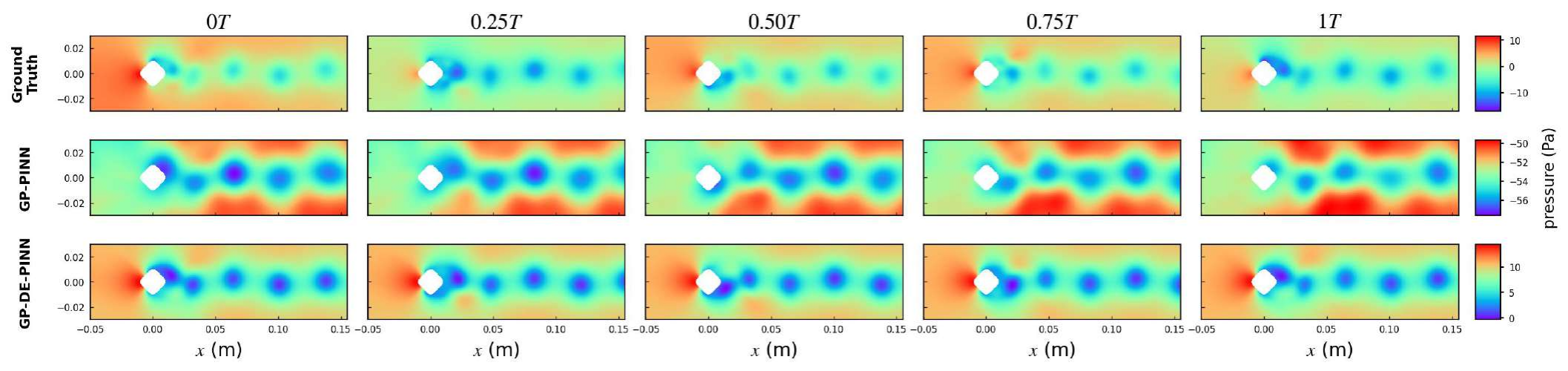}}
  \caption{\label{8-4pinn&gppinn} The flow fields around the petal-shaped cylinder ($r=8$ mm, $n=4$) in the test sets generated by the GP-PINN and GP-DE-PINN over the whole period, compared with the ground truth. (a) Contour of $u$. (b) Contour of $v$. (c) Contour of $p$. }
\end{figure}
\begin{figure}[!htb]
  \centering
  \subfloat[]{\includegraphics[width=1\textwidth]{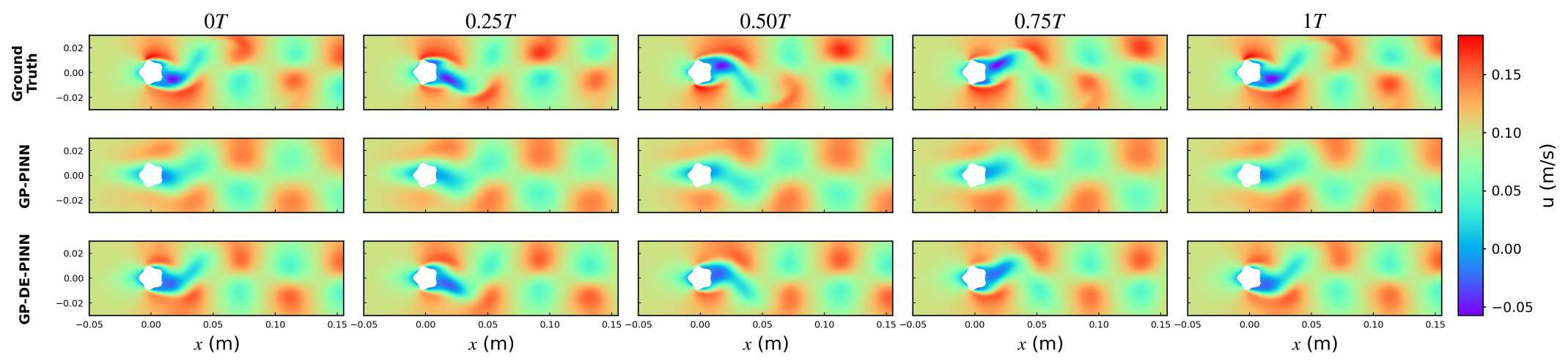}} \\
  \subfloat[]{\includegraphics[width=1\textwidth]{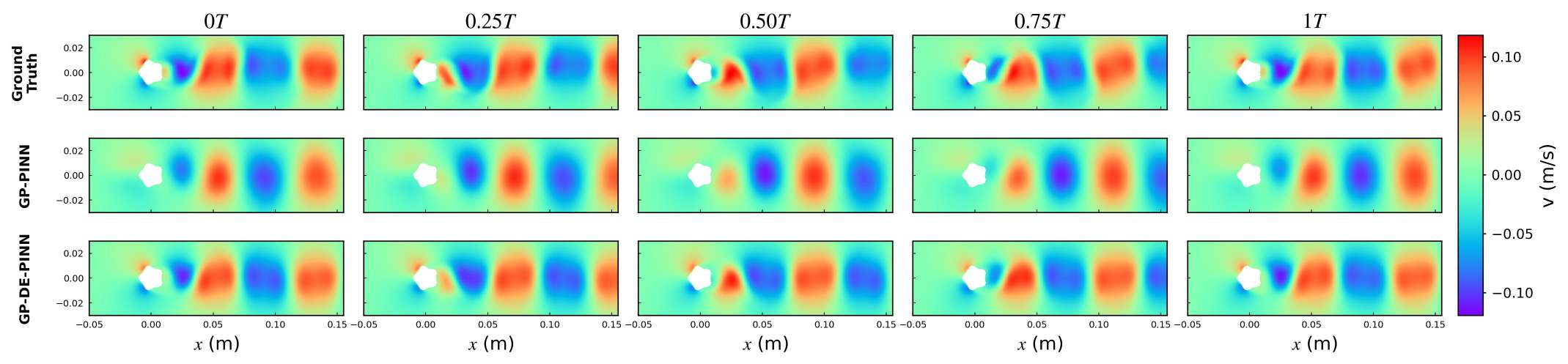}} \\
  \subfloat[]{\includegraphics[width=1\textwidth]{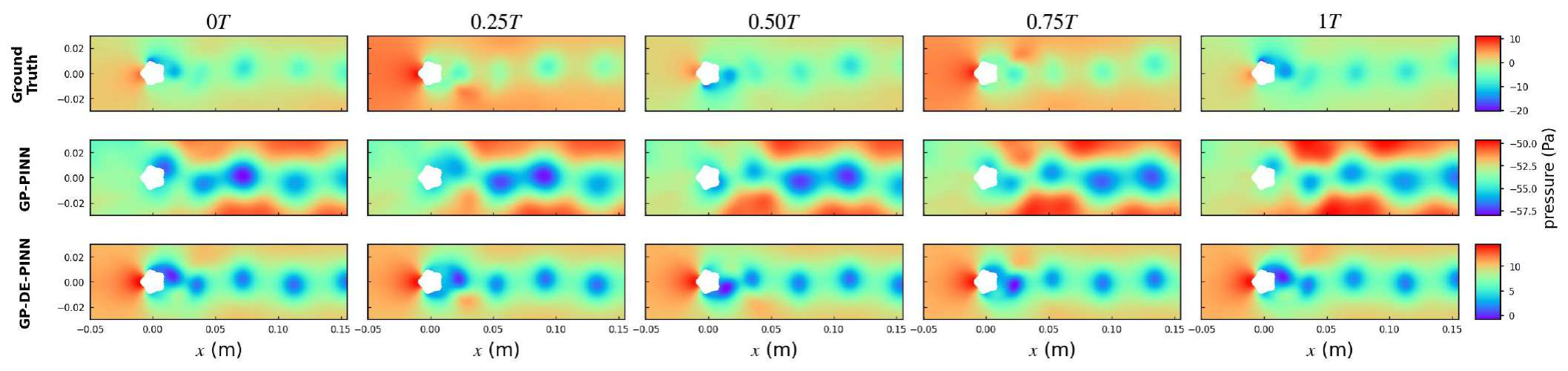}}
  \caption{\label{8-5pinn&gppinn} The flow fields around the petal-shaped cylinder ($r=8$ mm, $n=5$) in the test sets generated by the GP-PINN and GP-DE-PINN over the whole period, compared with the ground truth. (a) Contour of $u$. (b) Contour of $v$. (c) Contour of $p$. }
\end{figure}
\begin{figure}[!htb]
  \centering
  \subfloat[]{\includegraphics[width=1\textwidth]{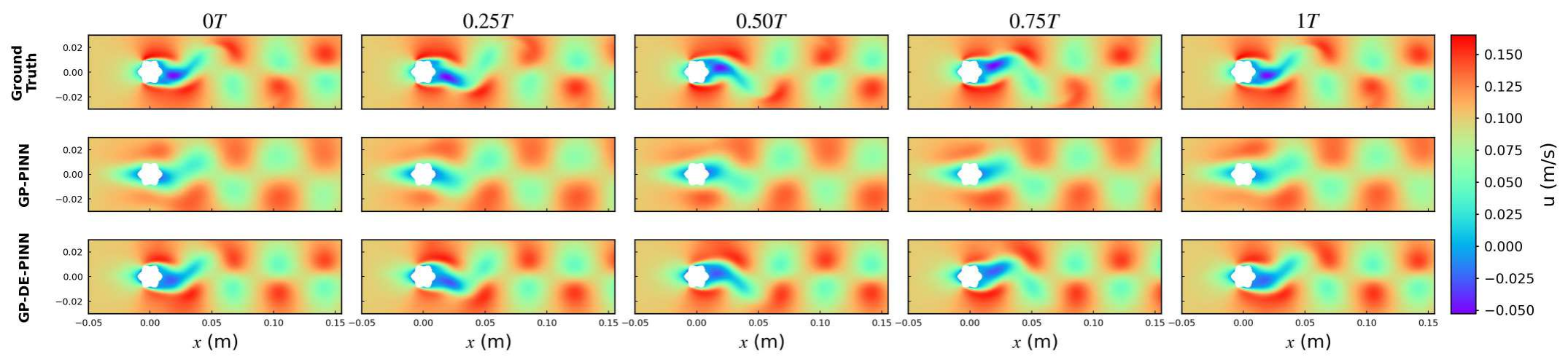}} \\
  \subfloat[]{\includegraphics[width=1\textwidth]{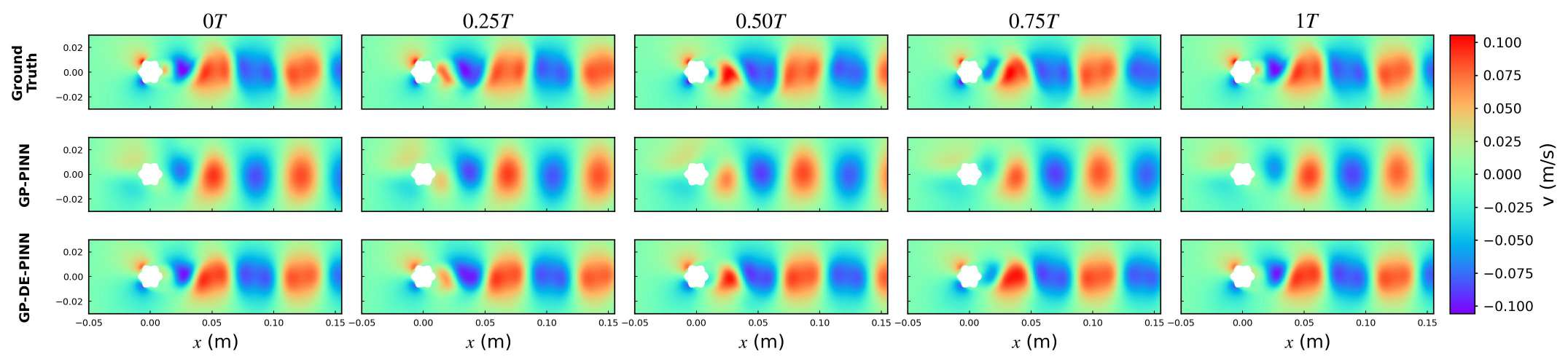}} \\
  \subfloat[]{\includegraphics[width=1\textwidth]{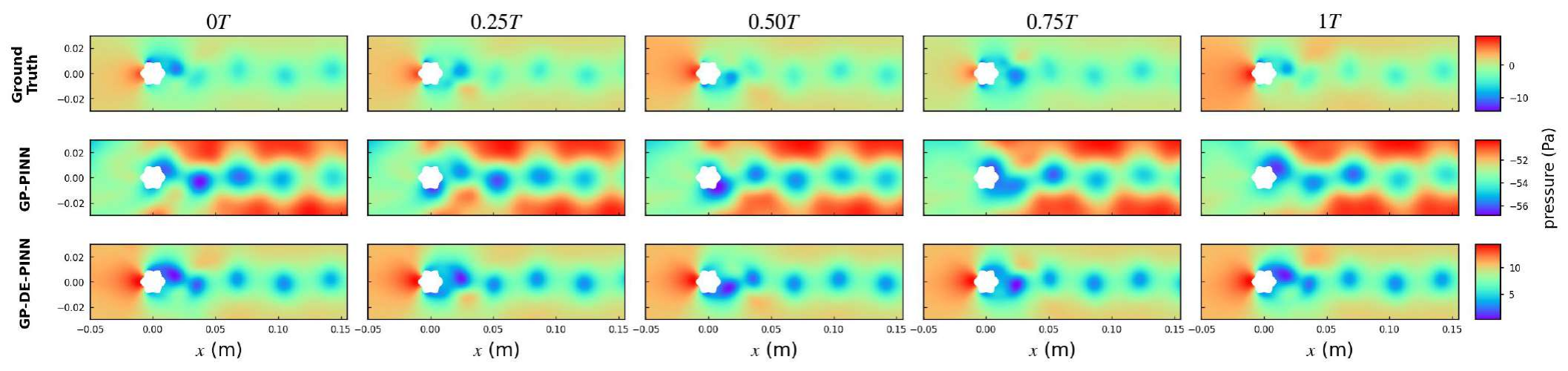}}
  \caption{\label{8-6pinn&gppinn} The flow fields around the petal-shaped cylinder ($r=8$ mm, $n=6$) in the test sets generated by the GP-PINN and GP-DE-PINN over the whole period, compared with the ground truth. (a) Contour of $u$. (b) Contour of $v$. (c) Contour of $p$. }
\end{figure}
\begin{figure}[!htb]
  \centering
  \subfloat[]{\includegraphics[width=1\textwidth]{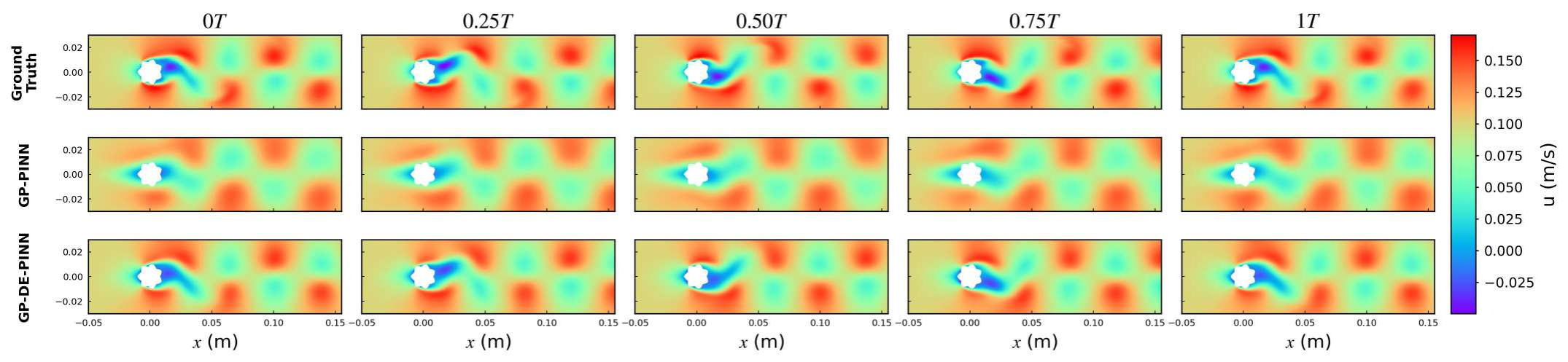}} \\
  \subfloat[]{\includegraphics[width=1\textwidth]{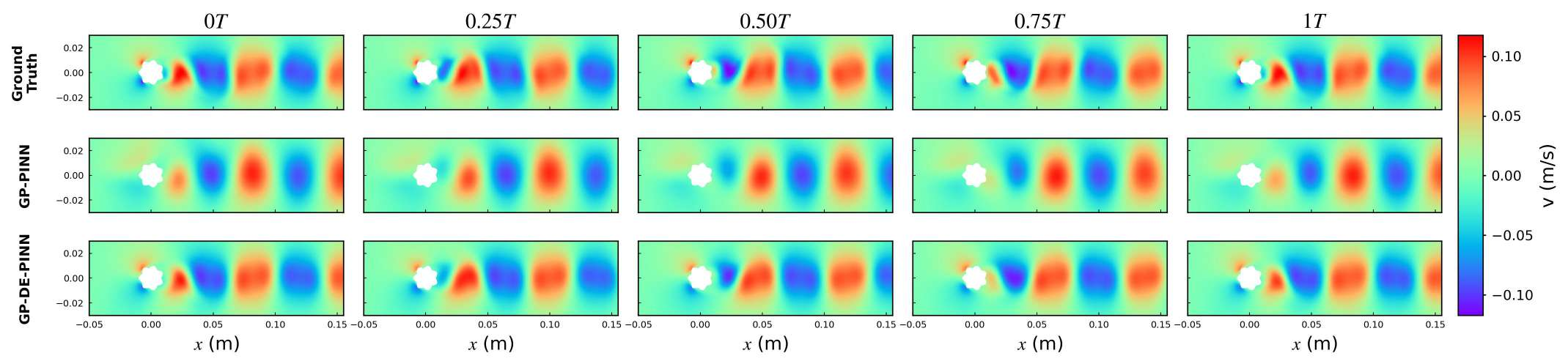}} \\
  \subfloat[]{\includegraphics[width=1\textwidth]{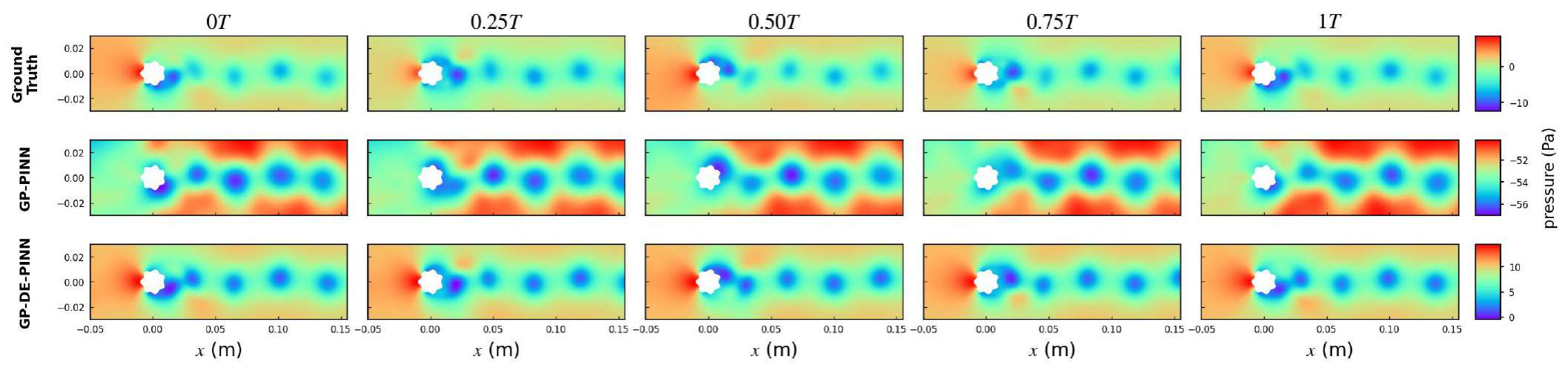}}
  \caption{\label{8-7pinn&gppinn} The flow fields around the petal-shaped cylinder ($r=8$ mm, $n=7$) in the test sets generated by the GP-PINN and GP-DE-PINN over the whole period, compared with the ground truth. (a) Contour of $u$. (b) Contour of $v$. (c) Contour of $p$. }
\end{figure}
\begin{figure}[!htb]
  \centering
  \subfloat[]{\includegraphics[width=1\textwidth]{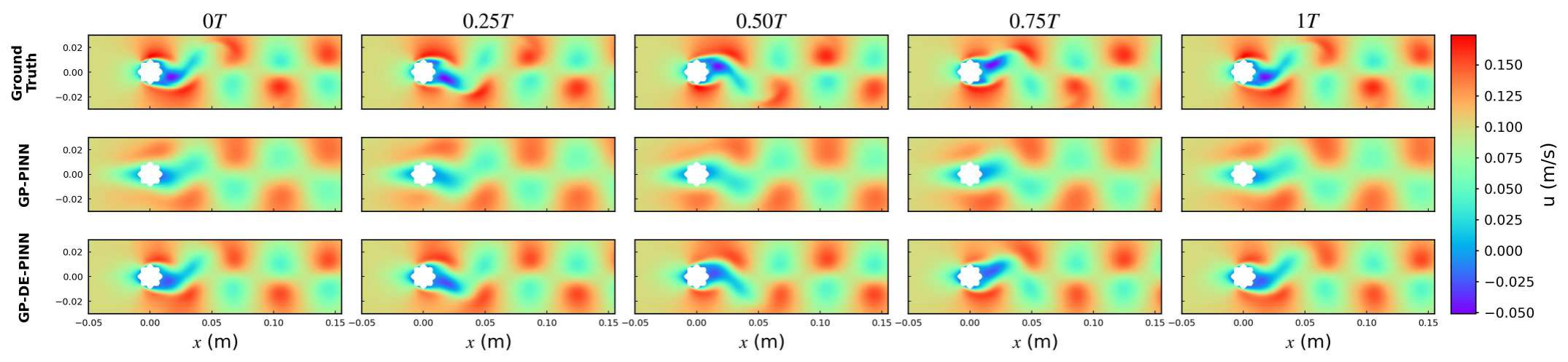}} \\
  \subfloat[]{\includegraphics[width=1\textwidth]{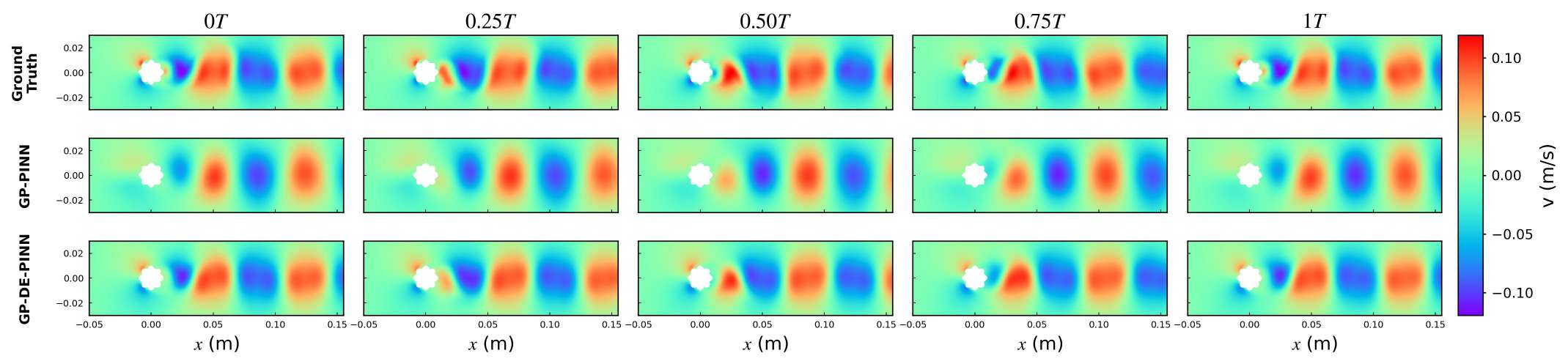}} \\
  \subfloat[]{\includegraphics[width=1\textwidth]{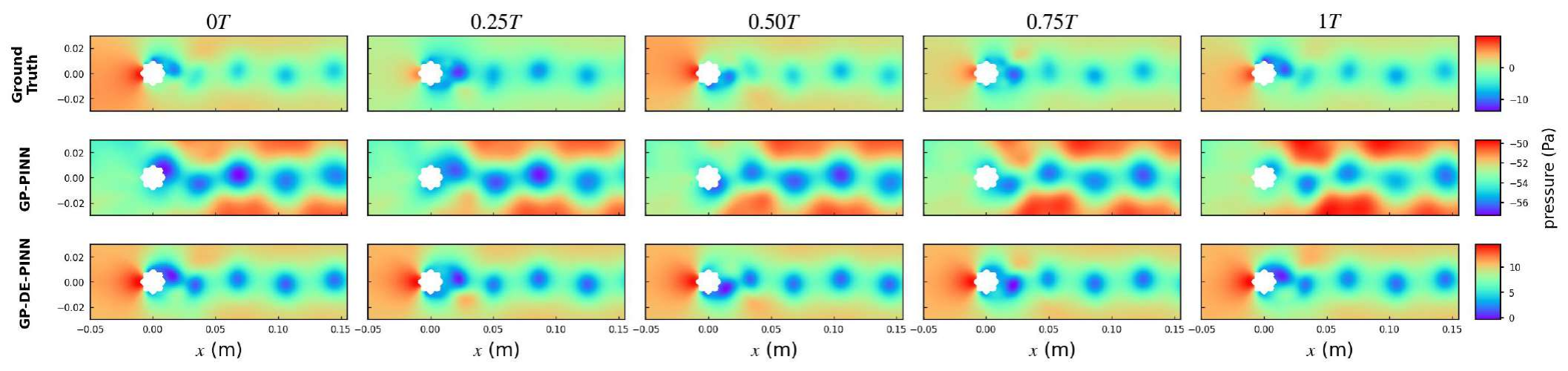}}
  \caption{\label{8-8pinn&gppinn} The flow fields around the petal-shaped cylinder ($r=8$ mm, $n=8$) in the test sets generated by the GP-PINN and GP-DE-PINN over the whole period, compared with the ground truth. (a) Contour of $u$. (b) Contour of $v$. (c) Contour of $p$. }
\end{figure}

\Cref{8-4pinn&gppinn} to \cref{8-8pinn&gppinn} visually present the spatiotemporal evolution of the velocity components ($u, v$) and the pressure field ($p$) across the test configurations. The comparison results show that GP-PINN performs poorly when predicting the flow fields for these new shapes, resulting in blurred shear layers and attenuated velocity amplitudes. In contrast, GP-DE-PINN demonstrates superior accuracy in all tested petal counts ($n=4 \sim 8$), precisely capturing the distinct shear layer separation and vortex topology.

Regarding the pressure field ($p$), consistent with the observations in the training cases, the GP-DE-PINN accurately predicts the trends of spatiotemporal evolution on these unseen geometries, despite the complete exclusion of pressure data from the training process. While minor numerical discrepancies in absolute magnitude persist due to the lack of direct supervision, the model correctly identifies high-pressure stagnation points and low-pressure vortex cores. This confirms that the GP-DE-PINN has successfully learned the intrinsic non-linear mapping between geometry and fluid dynamics, enabling it to infer the latent pressure variable via the Navier-Stokes constraints even for shapes outside the training set.

\Cref{tab:comparison} presents a detailed quantitative comparison of the prediction errors for the horizontal ($u$) and vertical ($v$) velocity components between GP-PINN and GP-DE-PINN for varying geometric parameters ($n=4$ to $n=8$). Predictive Performance is evaluated using the root mean square error (RMSE) and the mean relative error (MRE). In general, the tabulated data demonstrate that the GP-DE-PINN consistently outperforms the GP-PINN across all geometries examined.

For the horizontal velocity ($u$-component), the GP-PINN exhibits MRE values that generally fluctuate between 10\% and 16\%, with RMSE values hovering around 0.016. In contrast, the GP-DE-PINN achieves a substantial reduction in prediction error in all geometries. Specifically, for cases $n=6$ and $n=8$, the GP-DE-PINN demonstrates exceptional precision, suppressing the MRE to 4.42\% and 4.62\%, respectively. This represents an error reduction of more than 60\% compared to the GP-PINN, which records 11.11\% and 11.39\% for the same conditions. Even in the case of $n=5$, where errors are relatively higher for both models, GP-DE-PINN maintains the MRE at 10.22\%, significantly outperforming the GP-PINN's 15.91\%.

Predicting the vertical flow component ($v$-component) proves to be significantly more challenging, as evidenced by the GP-PINN's performance where MRE values exceed 25\% across all test cases, peaking at 35.74\% for $n=5$. However, the GP-DE-PINN achieves a breakthrough in stability with respect to this component. At $n=8$, GP-DE-PINN records an MRE of the $v$-component of only 9.46\%, while the GP-PINN remains high at 25.39\%. This highlights the GP-DE-PINN's capability to accurately capture complex flow features where the GP-PINN model fails. Furthermore, in other scenarios such as $n=4$ and $n=6$, the GP-DE-PINN model stabilizes the MRE at approximately 11\%, achieving a relative error reduction by an average of 57.3\% compared to the GP-PINN, effectively addressing the traditional weakness model in the prediction of vertical velocity. In summary, GP-DE-PINN provides predictions that are significantly closer to the ground truth.

\begin{table}[!htb]
\caption{\label{tab:comparison} Comparison of RMSE and MRE for $u$ and $v$ velocity components. GP-DE-PINN shows consistent improvement over GP-PINN across all geometries in the test set.}
\begin{ruledtabular}
\begin{tabular}{cccccc}
 & & \multicolumn{2}{c}{$u$-component} & \multicolumn{2}{c}{$v$-component} \\
 \cline{3-4} \cline{5-6}
 Geometry & Model & RMSE & MRE (\%) & RMSE & MRE (\%) \\
 \hline
 $n=4$
    & GP-PINN    & 0.016 & 10.81 & 0.017 & 26.15 \\
    & GP-DE-PINN & \textbf{0.008} & \textbf{5.49}  & \textbf{0.007} & \textbf{11.19} \\
 \hline
 $n=5$
    & GP-PINN    & 0.021 & 15.91 & 0.020 & 35.74 \\
    & GP-DE-PINN & \textbf{0.013} & \textbf{10.22} & \textbf{0.013} & \textbf{22.18} \\
 \hline
 $n=6$
    & GP-PINN    & 0.016 & 11.11 & 0.014 & 27.05 \\
    & GP-DE-PINN & \textbf{0.007} & \textbf{4.42}  & \textbf{0.006} & \textbf{11.51} \\
 \hline
 $n=7$
    & GP-PINN    & 0.015 & 10.30 & 0.014 & 25.06 \\
    & GP-DE-PINN & \textbf{0.008} & \textbf{5.51}  & \textbf{0.010} & \textbf{15.41} \\
 \hline
 $n=8$
    & GP-PINN    & 0.016 & 11.39 & 0.015 & 25.39 \\
    & GP-DE-PINN & \textbf{0.007} & \textbf{4.62}  & \textbf{0.006} & \textbf{9.46} \\
\end{tabular}
\end{ruledtabular}
\end{table}

\begin{figure}[!h]
  \centering
  \subfloat[\label{8-4avg&std_u}]{\includegraphics[width=0.495\textwidth]{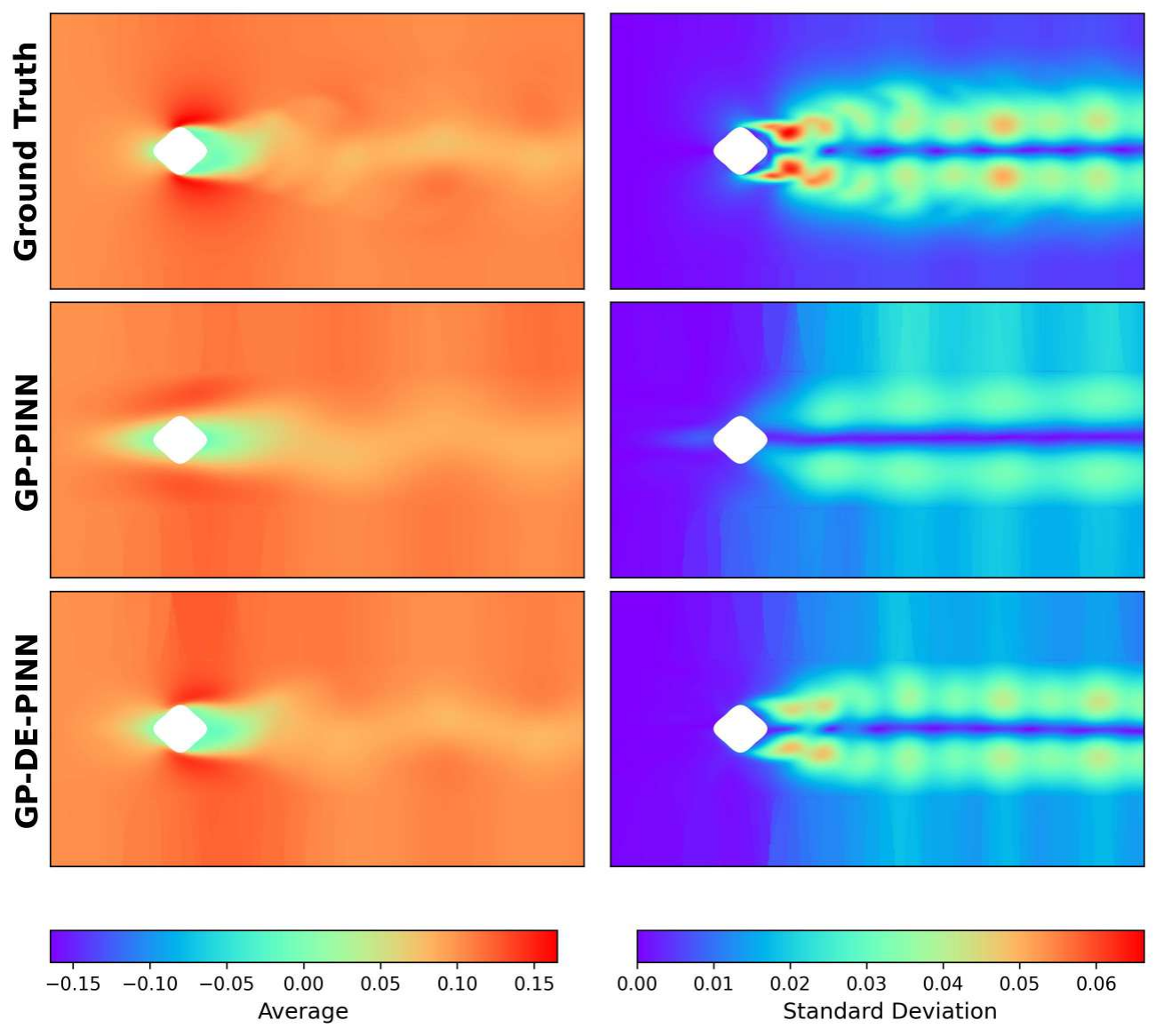}} 
  \subfloat[\label{8-4avg&std_v}]{\includegraphics[width=0.495\textwidth]{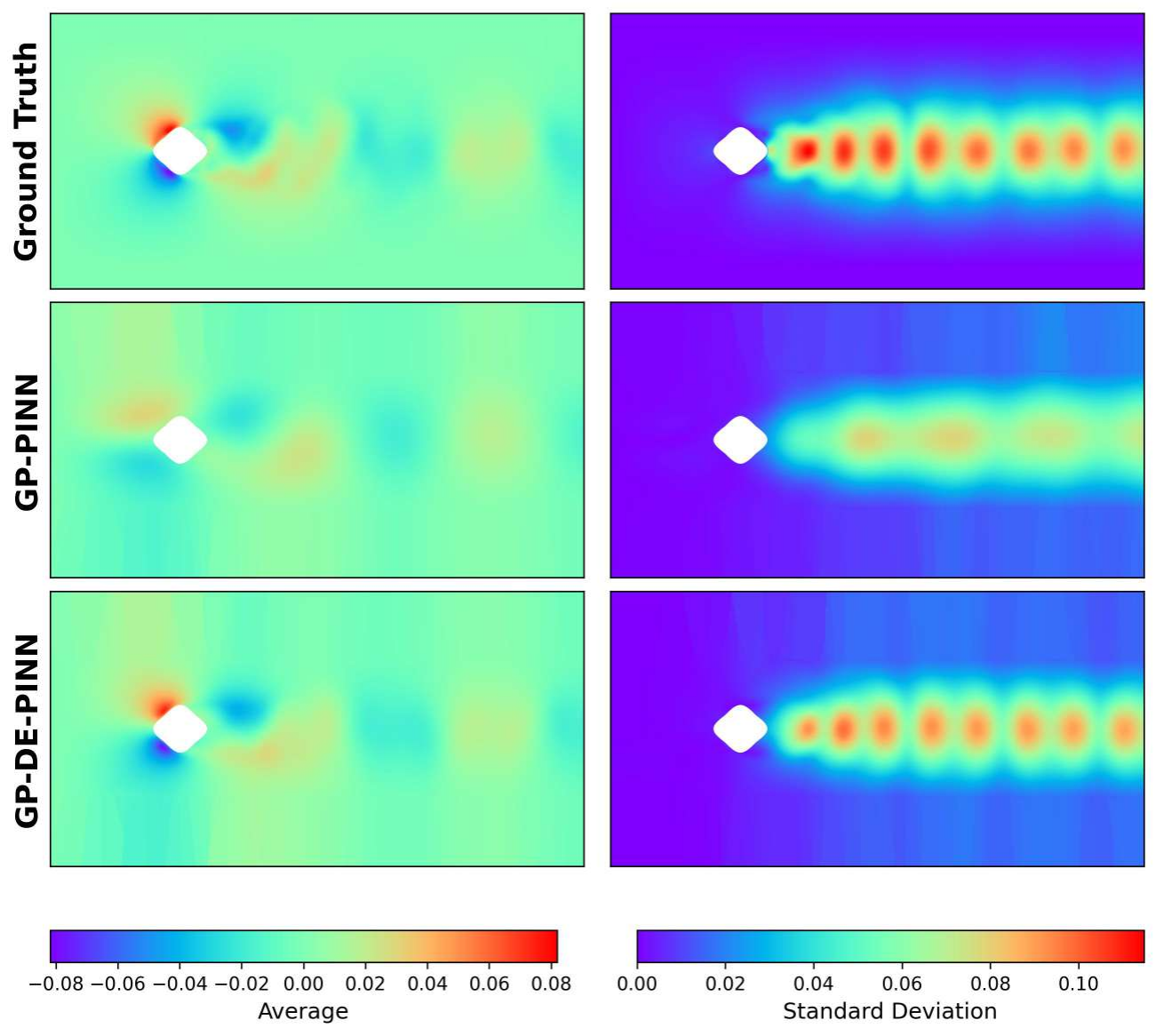}}
  \caption{\label{8-4avg&std} Statistical evaluation of GP-PINN and GP-DE-PINN on flow around the petal-shaped cylinder ($r=8$ mm, $n=4$) in the test sets. Statistics are computed over the whole period. (a) Mean and standard deviation of $u$. (b) Mean and standard deviation of $v$. }
\end{figure}

\begin{figure}[!h]
  \centering
  \subfloat[]{\includegraphics[width=0.495\textwidth]{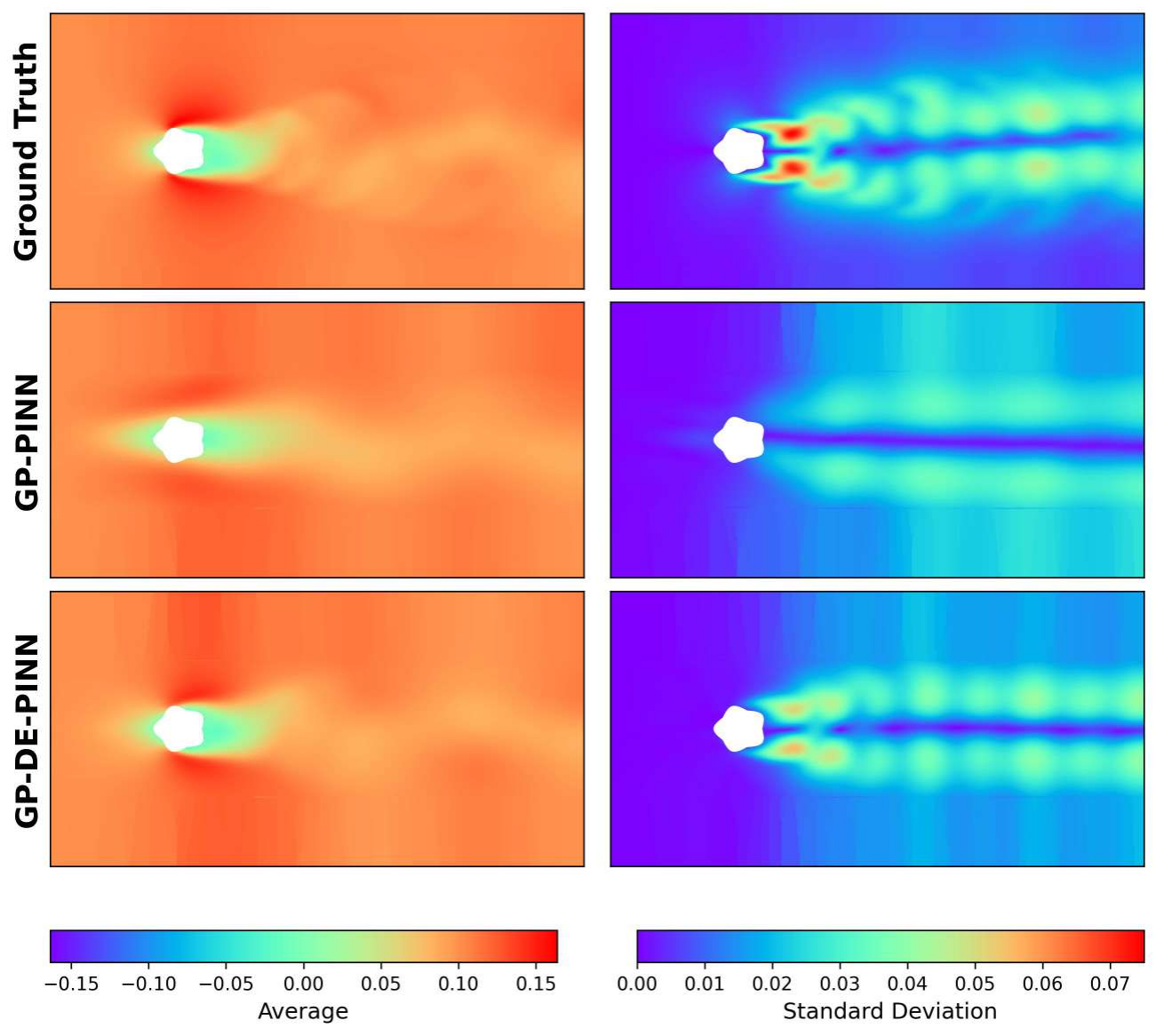}} 
  \subfloat[]{\includegraphics[width=0.495\textwidth]{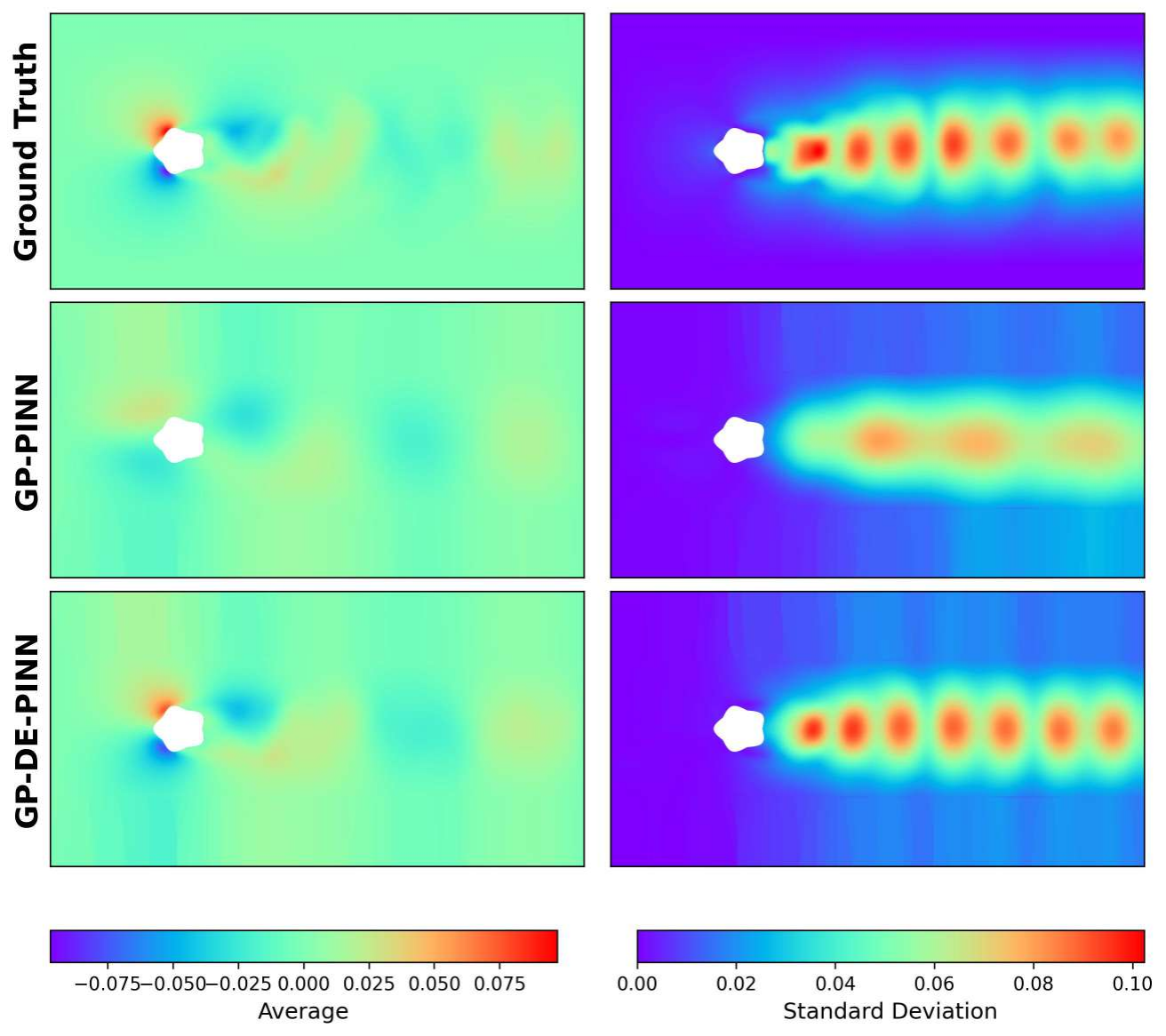}}
  \caption{\label{8-5avg&std} Statistical evaluation of GP-PINN and GP-DE-PINN on flow around the petal-shaped cylinder ($r=8$ mm, $n=5$) in the test sets. Statistics are computed over the whole period. (a) Mean and standard deviation of $u$. (b) Mean and standard deviation of $v$. }
\end{figure}

\begin{figure}[!h]
  \centering
  \subfloat[]{\includegraphics[width=0.495\textwidth]{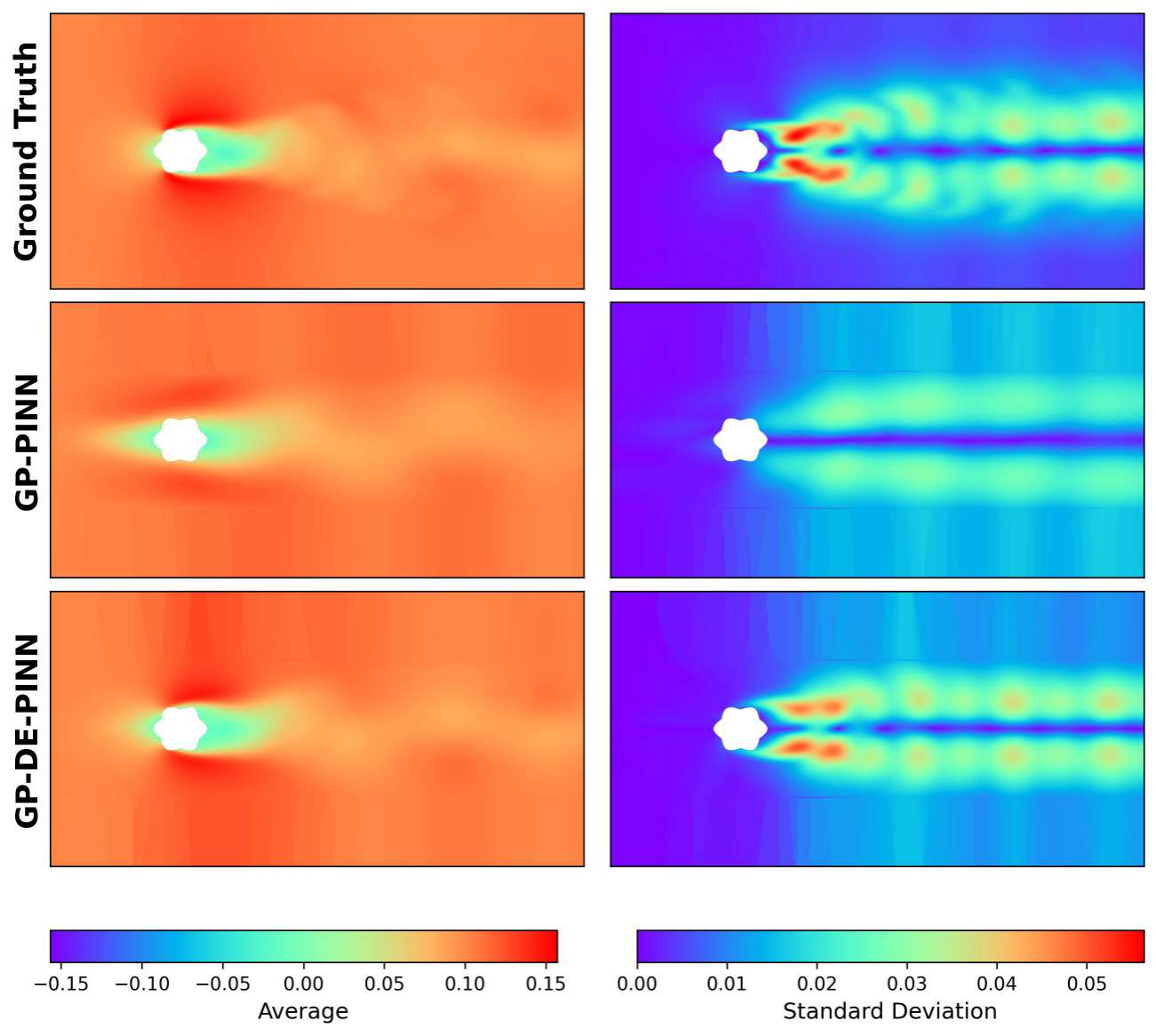}} 
  \subfloat[]{\includegraphics[width=0.495\textwidth]{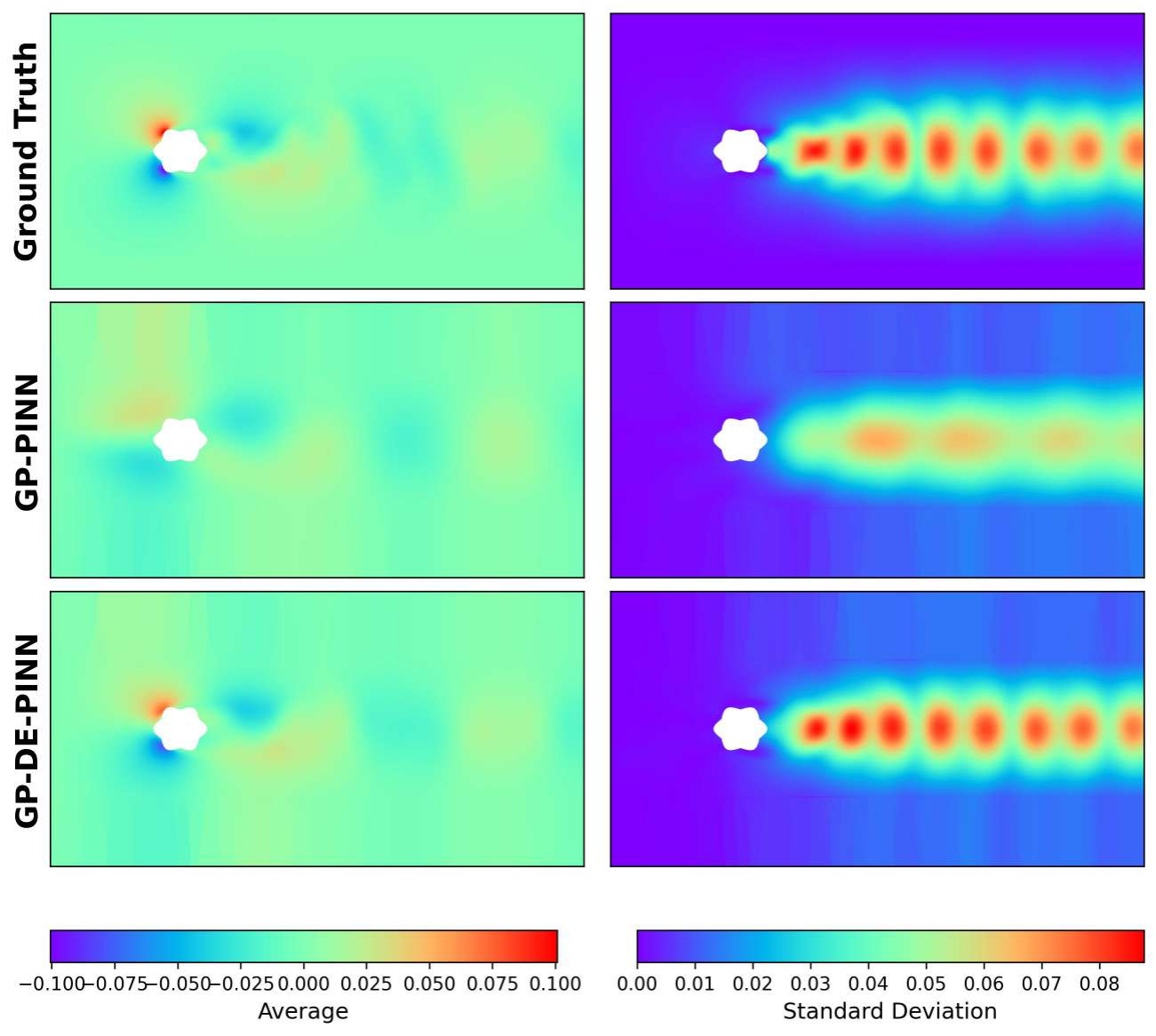}}
  \caption{\label{8-6avg&std} Statistical evaluation of GP-PINN and GP-DE-PINN on flow around the petal-shaped cylinder ($r=8$ mm, $n=6$) in the test sets. Statistics are computed over the whole period. (a) Mean and standard deviation of $u$. (b) Mean and standard deviation of $v$. }
\end{figure}

\begin{figure}[!h]
  \centering
  \subfloat[]{\includegraphics[width=0.495\textwidth]{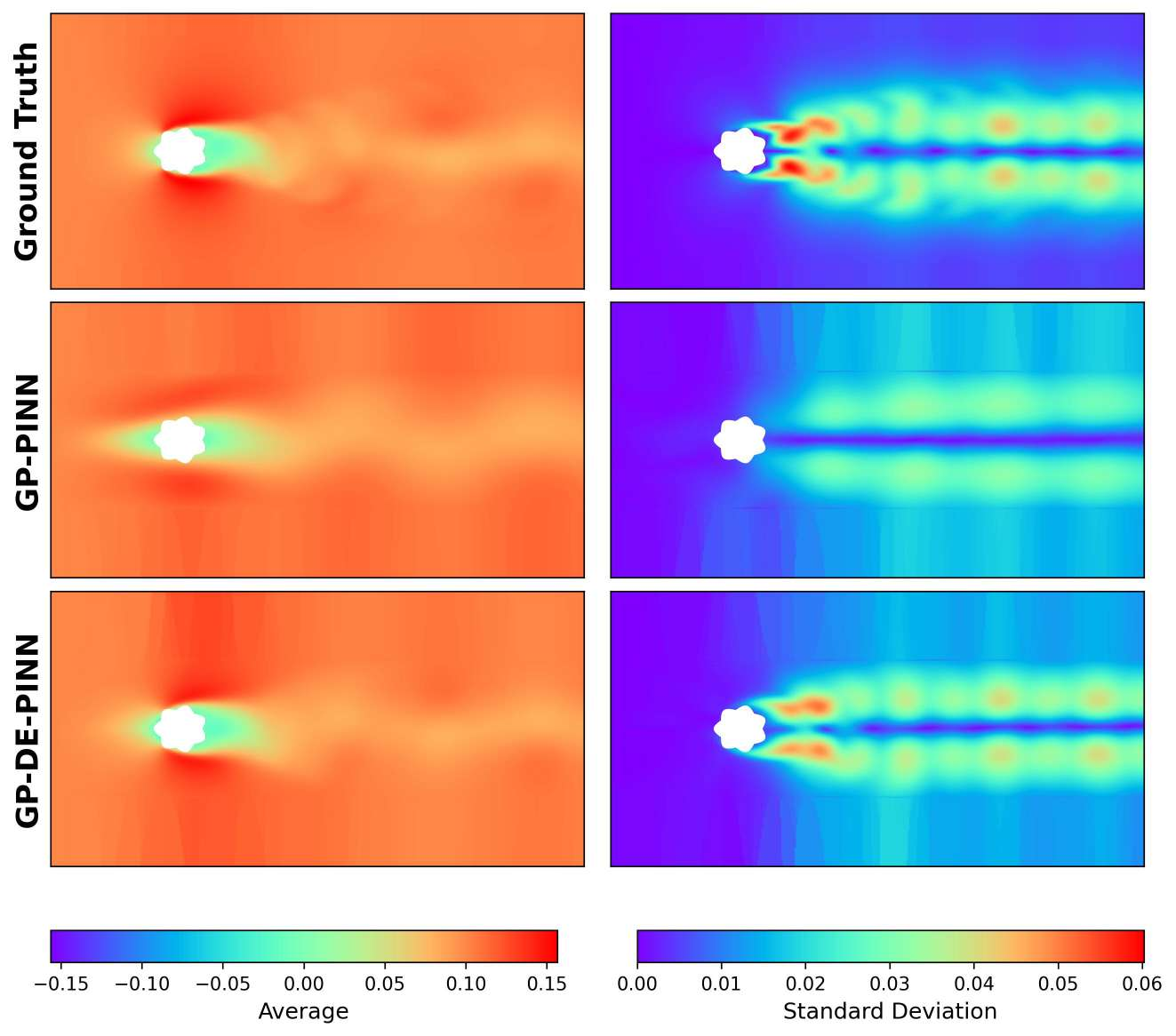}} 
  \subfloat[]{\includegraphics[width=0.495\textwidth]{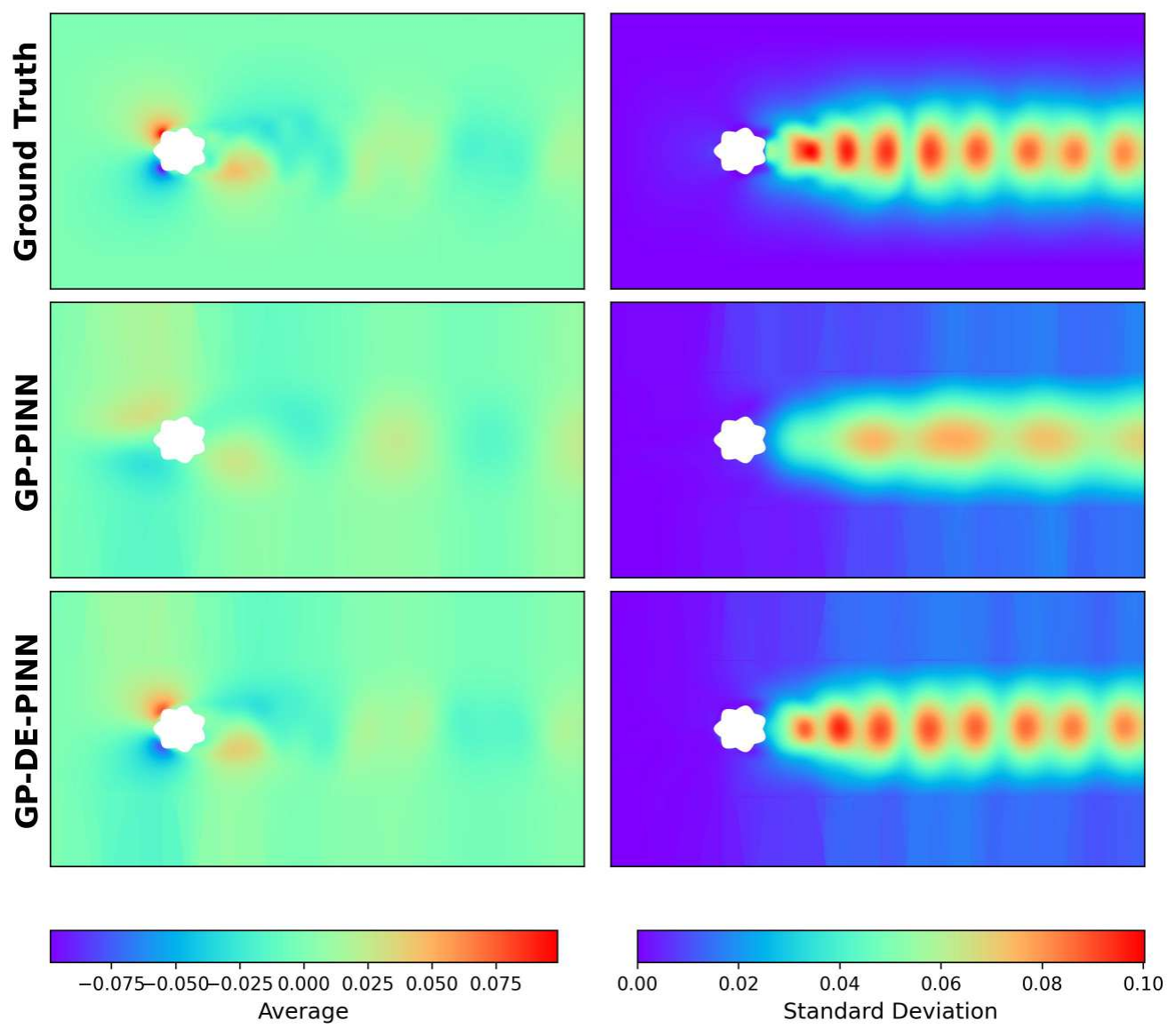}}
  \caption{\label{8-7avg&std} Statistical evaluation of GP-PINN and GP-DE-PINN on flow around the petal-shaped cylinder ($r=8$ mm, $n=7$) in the test sets. Statistics are computed over the whole period. (a) Mean and standard deviation of $u$. (b) Mean and standard deviation of $v$. }
\end{figure}

\begin{figure}[!h]
  \centering
  \subfloat[\label{8-8avg&std_u}]{\includegraphics[width=0.495\textwidth]{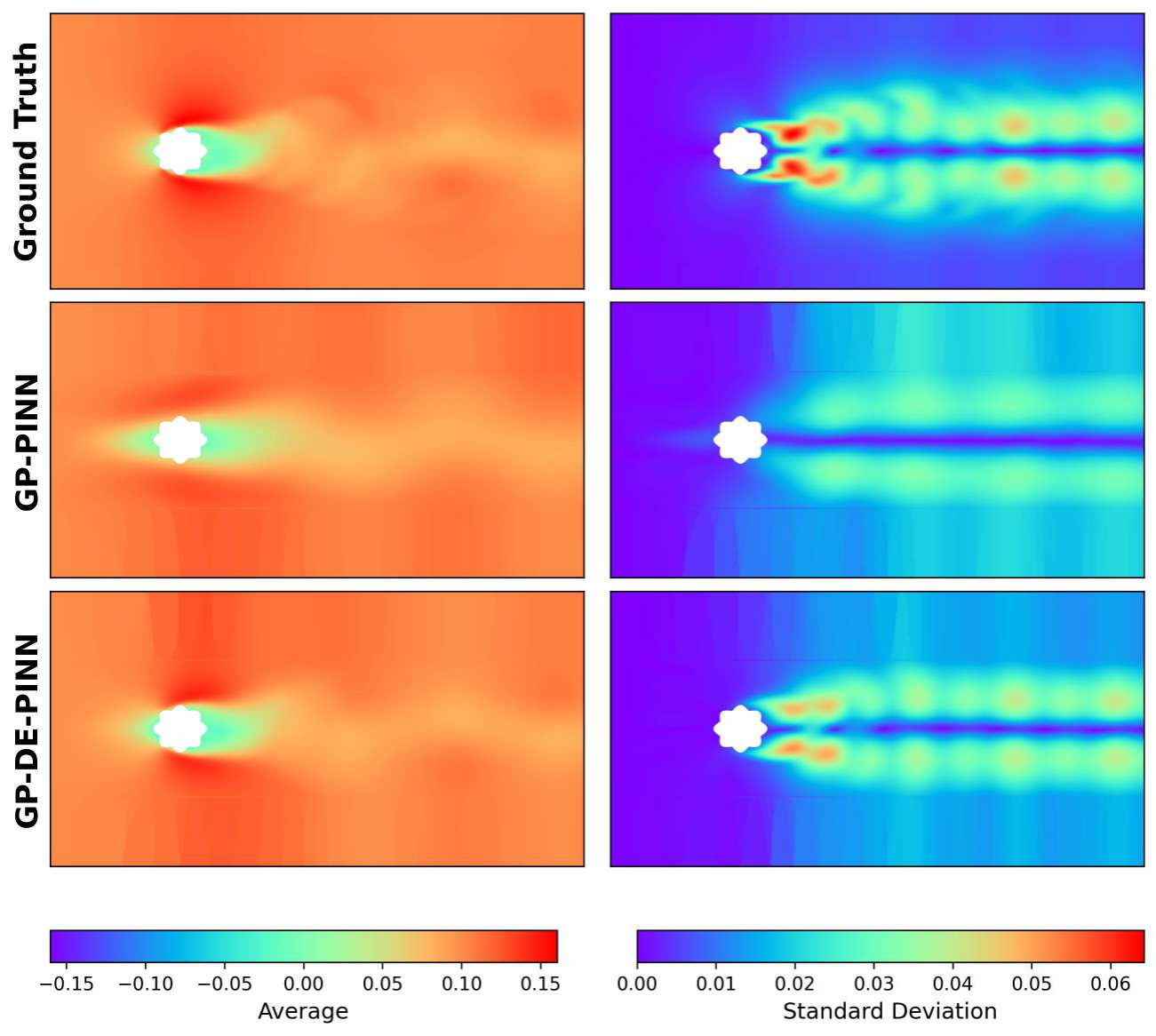}} 
  \subfloat[\label{8-8avg&std_v}]{\includegraphics[width=0.495\textwidth]{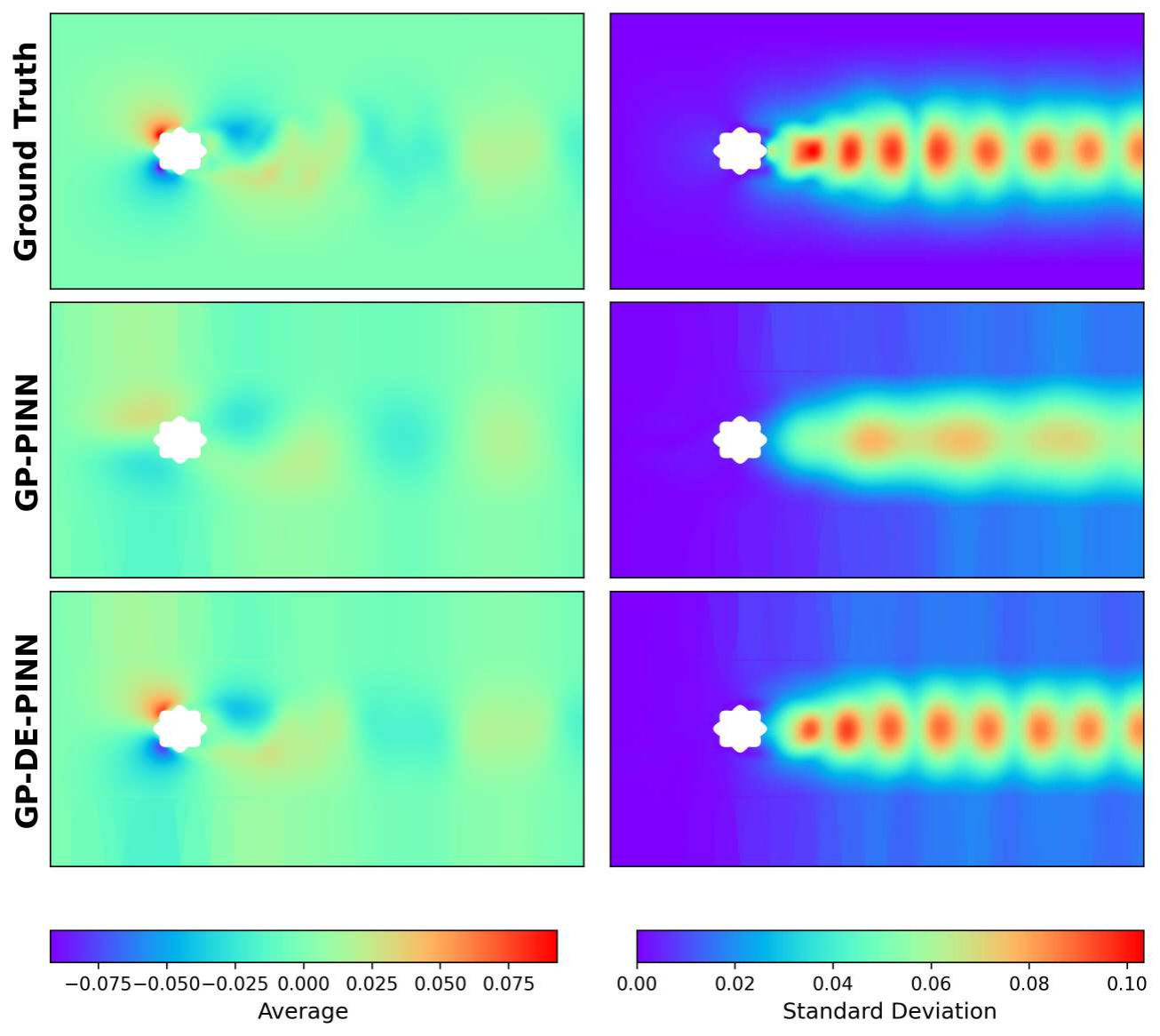}}
  \caption{\label{8-8avg&std} Statistical evaluation of GP-PINN and GP-DE-PINN on flow around the petal-shaped cylinder ($r=8$ mm, $n=8$) in the test sets. Statistics are computed over the whole period. (a) Mean and standard deviation of $u$. (b) Mean and standard deviation of $v$. }
\end{figure}

Same as the analysis pattern used for the training dataset, to provide a deeper insight into the physical accuracy of the reconstructed flow fields, we go beyond instantaneous snapshots and conducted a statistical analysis of the flow dynamics over a complete vortex shedding cycle ($T$). \Cref{8-4avg&std} to \cref{8-8avg&std} show the time-averaged mean and the standard deviation of the velocity components for unseen geometric configurations, such as the $r=8$ mm case with varying petal counts ($n$). These statistical metrics serve as critical indicators, where the mean field reveals the time-averaged wake structure, and the standard deviation quantifies the spatial distribution of the unsteady vortex street.

Regarding the time-averaged velocity fields, the results present a distinct low-velocity recirculation zone immediately downstream of the cylinder, characterized by a symmetric wake. However, the GP-PINN fails to preserve this correct wake characteristic. Conversely, the GP-DE-PINN accurately reproduces both the spatial extent and the magnitude of the mean wake. Across all tested petal configurations ($n=4$ to $8$), the contours of the time-averaged zero-velocity zone in the GP-DE-PINN predictions align precisely with the ground truth, demonstrating that the model has correctly captured the wake flow structures governed by the petal-shaped geometry.

Beyond the time-averaged velocity fields, the analysis of the standard deviation fields highlights the most significant disparity between the models. In the ground truth, high standard deviation values are concentrated along the shear layers and the path of the shedding vortex, forming distinct lobed structures that indicates the presence of strong periodic oscillations. The GP-PINN exhibits a critical failure in capturing these second-order statistics, predicting significantly attenuated fluctuation intensities. Especially in \cref{8-4avg&std_v} to \cref{8-8avg&std_v}, the GP-PINN model's deviation map has much lower peak values compared to the ground truth. In the contrast, the GP-DE-PINN maintains high-fidelity fluctuation statistics, successfully capturing the high-variance regions of $v$-component in the Karman vortex street. This ability to match the amplitude of the standard deviation confirms that the GP-DE-PINN preserves the characteristics of the flow.

The results above highlight GP-DE-PINN's ability to reconstruct and predict flow fields. Its dual-encoder architecture plays an important role in improving performance which transforms low-dimensional geometric parameters and spatiotemporal coordinates into high-dimensional latent feature vectors, enabling the inference of useful information about the obstacle's shape and flow field state.



\section{\label{sec:discussion}Discussion}

The preceding results have demonstrated the efficacy of the proposed GP-DE-PINN framework in accurately predicting flow fields around parametrically shaped cylinders. However, the model's predictive performance and generalization capability are intrinsically linked to the resolution of the geometric input representation and the expressive capacity of the neural architecture. To further evaluate the robustness of the framework and identify optimal hyperparameter configurations, this section conducts sensitivity analyses focusing on two critical aspects: the density of geometric parameter sampling and the network width of the geometric parameter encoder. By investigating the influence of the sampling interval $\Delta \theta$ and the number of hidden neurons $N_p$, we aim to clarify the relationship between these factors and prediction accuracy.

\subsection{Influence of geometric parameter sampling interval}

\begin{table}[!h]
\caption{\label{tab:comparison_theta} Comparison of prediction errors for different $\Delta \theta$ ($r=8$ mm), including group overall RMSE and MRE.}
\begin{ruledtabular}
\begin{tabular}{cccccccccc}
 & & \multicolumn{2}{c}{$u$-component} & \multicolumn{2}{c}{$v$-component} & \multicolumn{2}{c}{overall RMSE} & \multicolumn{2}{c}{overall MRE (\%)} \\
 \cline{3-4} \cline{5-6} \cline{7-8} \cline{9-10}
 $\Delta \theta$ & $n$ & RMSE & MRE (\%) & RMSE & MRE (\%) & $u$-comp. & $v$-comp. & $u$-comp. & $v$-comp. \\
 \hline
 \multirow{5}{*}{$5^\circ$} & 4 & $8.08 \times 10^{-3}$ & 5.18 & $7.50 \times 10^{-3}$ & 10.77 & \multirow{5}{*}{$8.07 \times 10^{-3}$} & \multirow{5}{*}{$7.39 \times 10^{-3}$} & \multirow{5}{*}{5.49} & \multirow{5}{*}{12.30} \\
 & 5 & $1.34 \times 10^{-2}$ & 9.64 & $1.17 \times 10^{-2}$ & 20.54 & & & & \\
 & 6 & $6.33 \times 10^{-3}$ & 4.13 & $5.41 \times 10^{-3}$ & 10.42 & & & & \\
 & 7 & $6.31 \times 10^{-3}$ & 4.12 & $5.96 \times 10^{-3}$ & 9.81 & & & & \\
 & 8 & $6.22 \times 10^{-3}$ & 4.39 & $6.39 \times 10^{-3}$ & 9.96 & & & & \\
 \hline
 \noalign{\smallskip}
 \multirow{5}{*}{$10^\circ$} & 4 & $8.59 \times 10^{-3}$ & 6.19 & $8.12 \times 10^{-3}$ & 12.59 & \multirow{5}{*}{$8.37 \times 10^{-3}$} & \multirow{5}{*}{$7.60 \times 10^{-3}$} & \multirow{5}{*}{6.12} & \multirow{5}{*}{13.07} \\
 & 5 & $1.30 \times 10^{-2}$ & 9.95 & $1.16 \times 10^{-2}$ & 20.80 & & & & \\
 & 6 & $6.82 \times 10^{-3}$ & 4.82 & $5.69 \times 10^{-3}$ & 11.24 & & & & \\
 & 7 & $6.43 \times 10^{-3}$ & 4.64 & $6.09 \times 10^{-3}$ & 10.21 & & & & \\
 & 8 & $7.02 \times 10^{-3}$ & 5.01 & $6.52 \times 10^{-3}$ & 10.50 & & & & \\
 \hline
 \noalign{\smallskip}
 \multirow{5}{*}{$20^\circ$} & 4 & $8.10 \times 10^{-3}$ & 5.49 & $7.37 \times 10^{-3}$ & 11.19 & \multirow{5}{*}{$8.60 \times 10^{-3}$} & \multirow{5}{*}{$8.37 \times 10^{-3}$} & \multirow{5}{*}{6.05} & \multirow{5}{*}{13.95} \\
 & 5 & $1.35 \times 10^{-2}$ & 10.22 & $1.27 \times 10^{-2}$ & 22.18 & & & & \\
 & 6 & $6.66 \times 10^{-3}$ & 4.42 & $5.77 \times 10^{-3}$ & 11.51 & & & & \\
 & 7 & $7.88 \times 10^{-3}$ & 5.51 & $9.92 \times 10^{-3}$ & 15.41 & & & & \\
 & 8 & $6.87 \times 10^{-3}$ & 4.62 & $6.08 \times 10^{-3}$ & 9.46 & & & & \\
 \hline
 \noalign{\smallskip}
 \multirow{5}{*}{$30^\circ$} & 4 & $1.21 \times 10^{-2}$ & 8.81 & $1.54 \times 10^{-2}$ & 23.12 & \multirow{5}{*}{$1.32 \times 10^{-2}$} & \multirow{5}{*}{$1.84 \times 10^{-2}$} & \multirow{5}{*}{10.01} & \multirow{5}{*}{29.16} \\
 & 5 & $1.67 \times 10^{-2}$ & 13.90 & $2.63 \times 10^{-2}$ & 42.50 & & & & \\
 & 6 & $1.31 \times 10^{-2}$ & 9.93 & $1.67 \times 10^{-2}$ & 29.98 & & & & \\
 & 7 & $1.40 \times 10^{-2}$ & 10.08 & $2.33 \times 10^{-2}$ & 34.01 & & & & \\
 & 8 & $1.00 \times 10^{-2}$ & 7.32 & $1.02 \times 10^{-2}$ & 16.19 & & & & \\
\end{tabular}
\end{ruledtabular}
\end{table}

The sampling interval $\Delta \theta$ of the geometric parameters directly determines the dimensionality of the input feature vector $\mathbf{d}$. To investigate the influence of geometric input density on the prediction accuracy of the GP-DE-PINN, we evaluate the model's performance across varying sampling intervals $\Delta \theta \in \{5^\circ, 10^\circ, 20^\circ, 30^\circ\}$. \Cref{tab:comparison_theta} presents a comparison of the RMSE and MRE for the predicted velocity components under different petal numbers $n$, along with the overall RMSE and MRE across all test cases.

A general trend is observed that the prediction error increases significantly when the sampling interval $\Delta \theta$ is large. As $\Delta \theta$ increases to $30^\circ$, the sparse representation of the geometry prevents the model from correctly learning the geometric features of the cylinder. This is quantitatively reflected in the overall MRE: when $\Delta \theta$ shifts from $20^\circ$ to $30^\circ$, the overall MRE for the $u$-component jumps from $6.05\%$ to $10.01\%$, and the $v$-component surges drastically from $13.95\%$ to $29.16\%$. Specifically, for the $n=5$ configuration, the $v$-component MRE effectively doubles from $20.54\%$ at $5^\circ$ to $42.50\%$ at $30^\circ$. These results show that an overly low sampling rate do not adequately capture critical local curvature details, significantly degrading predictive capability.

As the sampling interval $\Delta \theta$ decreases from $20^\circ$ to $5^\circ$ (i.e., the geometric input density increases), the downward trend in error metrics becomes less pronounced. the overall RMSE for the $u$-component steadily decreases from $8.60 \times 10^{-3}$ at $20^\circ$ to $8.07 \times 10^{-3}$ at $5^\circ$, while the $v$-component exhibits a similar decline from $8.37 \times 10^{-3}$ to $7.39 \times 10^{-3}$. Besides, the overall MRE for the $v$-component  drops from $13.95\%$ to $12.30\%$. The results suggest that for this specific set of petal-shaped cylinders, a sampling interval of $\Delta \theta = 20^\circ$ is sufficient for the network to effectively capture geometric variations,  as further densification yields insignificant performance enhancement.

\subsection{Influence of neuron number for geometric parameter encoder}

\begin{table}[!h]
\caption{\label{tab:neuron_comparison} Comparison of prediction errors for different neuron counts in the geometric parameter encoder $g_{\theta_{p}}$ (fixed $\Delta \theta = 20^\circ$, $r=8$ mm).}
\begin{ruledtabular}
\resizebox{\textwidth}{!}{
\begin{tabular}{cccccccccc}
 & & \multicolumn{2}{c}{$u$-component} & \multicolumn{2}{c}{$v$-component} & \multicolumn{2}{c}{ overall RMSE} & \multicolumn{2}{c}{overall MRE (\%)} \\
 \cline{3-4} \cline{5-6} \cline{7-8} \cline{9-10}
 $N_p$ & $n$ & RMSE & MRE (\%) & RMSE & MRE (\%) & $u$-comp. & $v$-comp. & $u$-comp. & $v$-comp. \\
 \hline
 \multirow{5}{*}{200} & 4 & $9.75 \times 10^{-3}$ & 7.03 & $1.13 \times 10^{-2}$ & 16.74 & \multirow{5}{*}{$9.36 \times 10^{-3}$} & \multirow{5}{*}{$1.05 \times 10^{-2}$} & \multirow{5}{*}{7.03} & \multirow{5}{*}{17.23} \\
 & 5 & $1.43 \times 10^{-2}$ & 11.16 & $1.54 \times 10^{-2}$ & 26.74 & & & & \\
 & 6 & $7.17 \times 10^{-3}$ & 5.06 & $7.10 \times 10^{-3}$ & 14.23 & & & & \\
 & 7 & $8.39 \times 10^{-3}$ & 6.66 & $1.14 \times 10^{-2}$ & 17.09 & & & & \\
 & 8 & $7.18 \times 10^{-3}$ & 5.22 & $7.17 \times 10^{-3}$ & 11.33 & & & & \\
 \hline
 \noalign{\smallskip}
 \multirow{5}{*}{250} & 4 & $8.10 \times 10^{-3}$ & 5.49 & $7.37 \times 10^{-3}$ & 11.19 & \multirow{5}{*}{$8.60 \times 10^{-3}$} & \multirow{5}{*}{$8.37 \times 10^{-3}$} & \multirow{5}{*}{6.05} & \multirow{5}{*}{13.95} \\
 & 5 & $1.35 \times 10^{-2}$ & 10.22 & $1.27 \times 10^{-2}$ & 22.18 & & & & \\
 & 6 & $6.66 \times 10^{-3}$ & 4.42 & $5.77 \times 10^{-3}$ & 11.51 & & & & \\
 & 7 & $7.88 \times 10^{-3}$ & 5.51 & $9.92 \times 10^{-3}$ & 15.41 & & & & \\
 & 8 & $6.87 \times 10^{-3}$ & 4.62 & $6.08 \times 10^{-3}$ & 9.46 & & & & \\
 \hline
 \noalign{\smallskip}
 \multirow{5}{*}{500} & 4 & $1.08 \times 10^{-2}$ & 7.86 & $1.09 \times 10^{-2}$ & 16.90 & \multirow{5}{*}{$1.07 \times 10^{-2}$} & \multirow{5}{*}{$1.03 \times 10^{-2}$} & \multirow{5}{*}{7.95} & \multirow{5}{*}{17.64} \\
 & 5 & $1.50 \times 10^{-2}$ & 11.61 & $1.28 \times 10^{-2}$ & 23.85 & & & & \\
 & 6 & $8.85 \times 10^{-3}$ & 6.27 & $7.67 \times 10^{-3}$ & 15.45 & & & & \\
 & 7 & $1.01 \times 10^{-2}$ & 7.58 & $1.26 \times 10^{-2}$ & 19.75 & & & & \\
 & 8 & $8.80 \times 10^{-3}$ & 6.45 & $7.46 \times 10^{-3}$ & 12.27 & & & & \\
\end{tabular}
}
\end{ruledtabular}
\end{table}

To evaluate the sensitivity of the GP-DE-PINN to network width of the geometric parameter encoder, a comparative analysis is conducted with the number of hidden layer neuron inside the $g_{\theta_p}$ ($N_p \in \{200, 250, 500\}$), while keeping the geometric sampling resolution fixed at $\Delta \theta = 20^\circ$. \Cref{tab:neuron_comparison} summarizes the prediction errors for these configurations, alongside the calculated overall RMSE and MRE across all the test cases.

The results demonstrate that the proposed framework exhibits robustness to variations in network width $N_p$ of the geometric parameter encoder, with the error metrics displaying a slight U-shaped trend. Specifically, the configuration with $N_p = 250$ achieves the optimal performance, yielding the lowest global error metrics. The overall RMSE reaches its minimum values of $8.60 \times 10^{-3}$ for the $u$-component and $8.37 \times 10^{-3}$ for the $v$-component, corresponding to the lowest overall MREs of $6.05\%$ and $13.95\%$, respectively. In comparison, the configuration of $N_p = 200$ shows slightly higher residuals, with overall RMSE values of $9.36 \times 10^{-3}$ ($u$) and $1.05 \times 10^{-2}$ ($v$). While the model with $N_p = 200$ already possesses adequate capacity to capture complex boundary geometric features, increasing the neuron count to $250$ further refines the feature extraction process, reducing the prediction error.

However, further expanding the network width $N_p$ from 250 to 500 does not yield continued improvement. Instead, a slight reduction  in performance is observed, as the overall RMSE rebounds to $1.07 \times 10^{-2}$ ($u$) and $1.03 \times 10^{-2}$ ($v$), accompanied by a rise in MRE to $7.95\%$ and $17.64\%$. This consistent degradation across both RMSE and MRE suggests that excessive parameterization may introduce optimization challenge and overfitting. While, the error levels at $N_p = 500$ remain comparable to those of the $N_p = 200$ case. This insensitivity, where deviations in the neuron count of $g_{\theta_p}$ do not lead to significant performance deterioration, indicates that the proposed geometry encoder effectively extracts latent feature vectors $h_{gp}$ without requiring precise hyperparameter tuning, proving its stability and reliability for practical applications.

\section{\label{sec:conclusion}Conclusion}

In this study, we propose the GP-DE-PINN, a unified framework designed to predict laminar flow fields around parametrically varying geometries. By integrating a specialized dual-encoder architecture consisting of a geometric parameter encoder and a spatiotemporal coordinate encoder, the model effectively integrates geometric features with spatiotemporal coordinates, enabling robust prediction across continuous boundary variations. We establish a parametric dataset based on petal-shaped geometries to evaluate the model’s interpolation and generalization capabilities, particularly on unseen geometric configurations.

The results demonstrate that the GP-DE-PINN significantly outperforms the GP-PINN (which incorporates geometric parameters as direct additional inputs) across both training samples and unseen test geometries. The GP-PINN tends to produce overly smooth and blurred flow fields, which leads to substantial deviations when predicting flow fields around unseen configurations, particularly resulting in high errors for the vertical velocity component. In contrast, the GP-DE-PINN exhibits high fidelity, reducing the relative error of velocity predictions by 50\% on average across different geometric configurations, compared to the GP-PINN. Notably, even without any pressure data during training, the GP-DE-PINN effectively infers the pressure relative distribution solely based on embedded physical equations, demonstrating its strong capability to deduce unknown fluid variables through physical laws.

Besides, the statistical analyses of flow predictions further illustrate the accuracy of the proposed method. The GP-DE-PINN accurately reproduces the time-averaged wake structures across various unseen geometric configurations. In terms of the standard deviation metric, the GP-PINN predicts significantly attenuated fluctuation intensities. In contrast, the GP-DE-PINN effectively resolves the second-order statistics, capturing the high-variance regions and preserving the strong fluctuation intensity associated with vortex shedding in the wake.

Additionally, sensitivity analyses regarding hyperparameters demonstrate the robustness of the GP-DE-PINN framework. For geometric parameter sampling, although the error decreases significantly when the sampling interval $\Delta \theta$ decreases from $30^\circ$ to $20^\circ$, the interval decreases from $20^\circ$ to $5^\circ$ only yields marginal error reduction. It indicates that for the dataset considered in this study, a sampling interval $\Delta \theta = 20^\circ$ is sufficient to represent geometric feature variations of the cylinders. Besides, the geometric parameter encoder exhibits a slight U-shaped error trend, optimizing accuracy at $N_p = 250$. The predictive performance of the model shows robustness against variations in  the network width of geometric parameter encoder.

In general, the proposed GP-DE-PINN demonstrates robust predictive capabilities for flow fields across varying geometric configurations. However, some extensions warrant further investigation. The current framework excludes pressure data during training. Future work will investigate the model performance when incorporating sparse pressure measurements. Besides, this work validates the model's ability to process different obstacle geometries. Future investigation will expand the dataset to encompass various Reynolds numbers and geometric variation. This will be achieved by integrating the Reynolds number as an additional input feature within the framework, thereby enhancing the model's generalization potential.

\begin{acknowledgments}
This work is supported by the Research Project of China COSCO Shipping Corporation Limited (2023-2-Z001-03), Open Project of the State Key Laboratory of Ocean Engineering, the Fundamental Research Funds for the Central Universities, the Independent Research Project of the State Key Laboratory of Ocean Engineering, and the Shanghai Jiao Tong University New Faculty Start-up Program.
\end{acknowledgments}

\section*{Data Availability Statement}
The data that support the findings of this study are available from the corresponding author upon reasonable request.

\bibliography{aipsamp}

\end{document}